\documentclass{svmult}
\usepackage{latexsym, amssymb, enumerate, amsmath, graphicx,subfig}
\input epsf

\usepackage {graphicx}

\begin{document}


\title*{Law of the Minimum Paradoxes}

\author{Alexander N. Gorban\inst{1} \and Lyudmila I. Pokidysheva\inst{2} \and Elena V. Smirnova\inst{2} \and Tatiana A.
Tyukina\inst{1}}
 \institute{Centre for Mathematical Modelling, University of Leicester,
Leicester, LE1 7RH,  UK,  \texttt{\{ag153,tt51\}@le.ac.uk} \and
Siberian Federal University, Krasnoyarsk, 660041, Russia
 }
 \authorrunning{A.N. Gorban \and L.I. Pokidysheva \and E.V. Smirnova \and T.A. Tyukina}
 \maketitle

\begin{abstract}
The ``Law of the Minimum" states that growth is controlled by the
scarcest resource (limiting factor). This concept was originally
applied to plant or crop growth (Justus von Liebig, 1840
\cite{Liebig1}) and quantitatively supported by many experiments.
Some generalizations based on more complicated ``dose-response"
curves were proposed. Violations of this law in natural and
experimental ecosystems were also reported. We study models of
adaptation in ensembles of similar organisms under load of
environmental factors and prove that violation of Liebig's law
follows from adaptation effects. If the fitness of an organism in a
fixed environment satisfies the Law of the Minimum then adaptation
equalizes the pressure of essential factors and therefore acts
against the Liebig's law. This is the {\it the Law of the Minimum
paradox}: if for a randomly chosen pair ``organism--environment" the
Law of the Minimum typically holds, then, in a well-adapted system,
we have to expect violations of this law.

For the opposite interaction of factors (a synergistic system of
factors which amplify each other) adaptation leads from factor
equivalence to limitations by a smaller number of factors.

For analysis of adaptation we develop a system of models based on
Selye's idea of the universal adaptation resource (adaptation
energy). These models predict that under the load of an
environmental factor a population separates into two groups
(phases): a less correlated, well adapted group and a highly
correlated group with a larger variance of attributes, which
experiences problems with adaptation. Some empirical data are
presented and evidences of interdisciplinary applications to
econometrics are discussed.
\end{abstract}


\section{Introduction}

\subsection{The Law of the Minimum}

The ``Law of the Minimum" states that growth is controlled by the
scarcest resource (limiting factor) \cite{Liebig1}. This law is
usually believed to be the result of  Justus von Liebig's research
(1840) but the agronomist and chemist Carl Sprengel published in
1828 an article that contained in essence the Law of the Minimum and
this law can be called the Sprengel--Liebig Law of the Minimum.
\cite{van der Ploeg1999}.

\begin{figure} \centering{
\includegraphics[width=50mm]{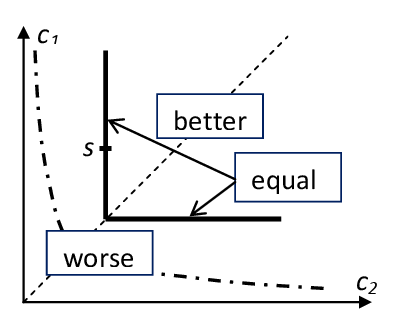}}
\caption{\label{Liebig} The Law of the Minimum. Coordinates $c_1$,
$c_2$ are normalized values of factors. For a given state
$s=(c_1(s),c_2(s))$, the bold solid line
$\min\{c_1,c_2\}=\min\{c_1(s),c_2(s)\}$ separates the states with
better conditions (higher productivity) from the states with worse
conditions. On this line the conditions do not differ significantly
from $s$ because of the same value of the limiting factor. The dot
dash line shows the border of survival. On the dashed line the
factors are equally important ($c_1=c_2$).}
\end{figure}

This concept is illustrated in Fig.~\ref{Liebig}.

This concept was originally applied to plant or crop growth. Many
times it was criticized, rejected, and then returned to and
demonstrated quantitative agreement with experiments
\cite{Liebig1,Liebig2+,Liebig3+}.

The Law of the Minimum was extended to a more general conception of
factors, rather than for the elementary physical description of
available chemical substances and energy. Any environmental factor
essential for life that is below the critical minimum, or that
exceeds the maximum tolerable level could be considered as a
limiting one.

There were several attempts to create a general theory of factors
and limitation in ecology, physiology and evolutionary biology.
Tilman \cite{Tilman1980} proposed an equilibrium theory of resource
competition based on classification of interaction in pairs of
resources. They may be: (1) essential, (2) hemi-essential, (3)
complementary, (4) perfectly substitutable, (5) antagonistic, or (6)
switching. This interaction depends on spatial heterogeneity of
resource distributions. For various resource types, the general
criterion for stable coexistence of species was developed.

Bloom, Chapin and Mooney
\cite{BloomChapinMooney1985,ChapinSchulzeMooney1990} elaborated the
economical metaphor of ecological concurrency. This analogy allowed
them to merge the optimality and the limiting approach and to
formulate four ``theorems". In particular, Theorem 3 states that a
plant should adjust allocation so that, for a given expenditure in
acquiring each resource, it achieves the same growth response:
growth is equally limited by all resources. This is a result of
adjustment: adaptation makes the limiting factors equally important.
They also studied the possibility for resources to substitute for
one another (Theorem 4) and introduced the concept of ``exchange
rate".

For human physiology the observation that adaptation makes the
limiting factors equally important was supported by many data of
human adaptation to the Far North conditions (or, which is the same,
disadaptation causes inequality of factors and leads to appearance
of single limiting factor) \cite{GorSmiCorAd1st}. The theory of
factors -- resource interaction was developed and supported by
experimental data. The results are used for monitoring of human
populations in Far North \cite{Sedov}.

In their perspectives paper, Sih and Gleeson \cite{SihGleeson1995}
considered three inter-related issues which form the core of
evolutionary ecology: (1) key environmental factors; (2) organismal
traits that are responses to the key factors; (3) the evolution of
these key traits. They suggested to focus on 'limiting traits'
rather than optimal traits. Adaptation leads to optimality and
equality of traits as well as of factors but under variations some
traits should be more limiting than others. From the Sih and Gleeson
point of view, there is a growing awareness of the potential value
of the limiting traits approach as a guide for studies in both basic
and applied ecology.

The critics of the Law of the Minimum is usually based on the
``colimitation" phenomenon: limitation of growth and survival by a
group of equally important factors and traits. For example, analysis
of species-specific growth and mortality of juvenile trees at
several contrasting sites suggests that light and other resources
can be simultaneously limiting, and challenges the application of
the Law of the Minimum to tree sapling growth \cite{Kobe1996}.

The concept of multiple limitation was proposed for unicellular
organisms based on the idea of the nutritional status of an organism
expressed in terms of state variables \cite{vandenBerg1998}. The
property of being limiting was defined in terms of the reserve
surplus variables. This approach was illustrated by numerical
experiments.

In the world oceans there are High Nutrient--Low Chlorophyll regions
where chlorophyll concentrations are lower than expected
concentrations given the ambient phosphate and nitrate levels. In
these regions, limitations of phytoplankton growth by other
nutrients like silicate or iron have been hypothesized and supported
by experiments. This colimitation was studied using a nine-component
ecosystem model embedded in the HAMOCC5 model of the oceanic carbon
cycle \cite{Aumontatal2003}.

The double--nutrient--limited growth appears also as a transition
regime between two regimes with single limiting factor. For bacteria
and yeasts at a constant dilution rate in the chemostat, three
distinct growth regimes were recognized: (1) a clearly
carbon-limited regime with the nitrogen source in excess, (2) a
double--nutrient--limited growth regime where both the carbon and
the nitrogen source were below the detection limit, and (3) a
clearly nitrogen-limited growth regime with the carbon source in
excess. The position of the double--nutrient--limited zone is very
narrow at high growth rates and becomes broader during slow growth
\cite{EgliZinn2003,ZinnWitholtEgli2004}.

Decomposition of soil organic matter is limited by both the
available substrate and the active decomposer community. The
colimitation effects strongly affect the feedbacks of soil carbon to
global warming and its consequences \cite{WutzlerReichstein2008}.

Dynamics of communities lead to colimitation on community level even
if organisms and populations remain limited by single factors.
Communities are likely to adjust their stoichiometry by competitive
exclusion and coexistence mechanisms. It guaranties simultaneous
limitation by many resources and optimal use of them at the
community scale. This conclusion was supported by a simple resource
ratio model and an experimental test carried out in microcosms with
bacteria \cite{Dangeratal2008}.

In spite of the long previous discussion of colimitation, in 2008
Saito and Goepfert stressed that this notion is ``an important yet
often misunderstood concept" \cite{SaitoGoepfert2008}. They describe
the potential nutrient colimitation pairs in the marine environment
and define three types of colimitation:
\begin{enumerate}\renewcommand{\labelenumi}{\Roman{enumi}.}
\item{{\em Independent nutrient colimitation} concerns two elements that
are generally biochemically mutually exclusive, but are also both
found in such low concentrations as to be potentially limiting.
Example: nitrogen--phosphorus colimitation.}
 \item{{\em Biochemical
substitution colimitation} involves two elements that can substitute
for the same biochemical role within the organism. Example:
zinc--cobalt colimitation.}
 \item{{\em Biochemically dependent colimitation} refers to the limitation of
one element that manifests itself in an inability to acquire another
element. Example: zinc--carbon colimitation.}
\end{enumerate}
\renewcommand{\labelenumi}{\arabic{enumi}.}

The experimental colimitation examples of the first type do not
refute the Law of the Minimum completely but rather support the
following statement: the ecological systems of various levels, from
an organism to a community, may avoid the monolimitation regime
either by the natural adjustment of their consumption structure
\cite{BloomChapinMooney1985,SemSem} or just by living in the
transition zone between the monolimitation regimes. From the general
point of view \cite{SihGleeson1995}, such a transition zone is
expected to be quite narrow (as a vicinity of a surface where
factors are equal) but in some specific situations it may be broad,
for example, for slow growth regimes in the chemostat
\cite{EgliZinn2003,ZinnWitholtEgli2004}.

The type II and type III colimitations should be carefully separated
from the usual discussion of the Law of the Minimum limitation. For
these types of colimitation, two (or more) nutrients limit growth
rates simultaneously, either through the effect of biochemical
substitution (type II) or by depressing the ability for the uptake
of another nutrient (type III) \cite{SaitoGoepfert2008}. The type II
and type III colimitations give us examples of the ``non-Liebig"
organization of the system of factors.

The Law of the Minimum is one of the most important tools for
mathematical modeling of ecological systems. It gives a clue for
constructing the first model for multi-component and multi-factor
systems. This clue sounds rather simple: first of all, we have to
take into account the most important factors which are, probably,
limiting factors. Everything else should be excluded and allowed
back only in a case when a ``sufficient reason" is proved (following
the famous ``Principle of Sufficient Reason" by  Leibnitz, one of
the four recognized laws of thought).

It is suggested to consider the Liebig production function as the
``archetype" for ecological modeling \cite{Nijlandatal2008}. The
generalizations of the Law of the Minimum were supported by the
biochemical idea of limiting reaction steps (see, for example,
\cite{Brown} or recent review \cite{GorRadLim}). Three classical
production functions, the Liebig, Mitscherlich and Liebscher
relations between nutrient supply and crop production, are limiting
cases of an integrated model based on the Michaelis--Menten kinetic
equation \cite{Nijlandatal2008}.

Applications of the Law of the Minimum to the ecological modeling
are very broad. The quantitative theories of the bottom--up control
of the phytoplankton dynamics is based on the influence of limiting
nutrients on growth and reproduction. The most used is the Droop
model and its generalizations
\cite{Droop1973,LegovicCruzado1997,Ballantyneatal2008}.

The Law of the Minimum was combined with the evolutionary dynamics
to analyze  the ``Paradox of the plankton" \cite{Shoreshatal2008}
formulated by Hutchinson \cite{Hutchinson1961} in 1961: ``How it is
possible for a number of species to coexist in a relatively
isotropic or unstructured environment all competing for the same
sorts of materials... According to the principle of competitive
exclusion... we should expect that one species alone would
outcompete all of the others." It was shown that evolution
exacerbates the paradox and it is now very far from the resolution.

The theory of evolution from monolimitation toward colimitation was
developed that takes into account the viruses attacks on the
phytoplankton receptors \cite{Menge2009}. In the classic theory
\cite{Tilman1982}, evolution toward colimitation decreases
equilibrium resource concentrations and increases equilibrium
population density. In contrary, under influence of viruses,
evolution toward co-limitation may have no effect on equilibrium
resource concentrations and may decreases the equilibrium population
density \cite{Menge2009}.

The Law of the Minimum was used for modeling of microcolonial fungi
growth on rock surfaces \cite{Chertovatal2004}. The analysis
demonstrated, that a continued lack of organic nutrition is a
dominating environmental factor limiting growth on stone monuments
and other exposed rock surfaces in European temperate and
Mediterranean climate.

McGill \cite{McGill2005} developed a model of coevolution of
mutualisms where one resource is traded for another resource. The
mechanism is based on the Law of the Minimum in combination with
Tilman's  approach to resource competition
\cite{Tilman1980,Tilman1982}. It was shown that resource limitations
cause mutualisms to have stable population dynamics.

The Law of the Minimum produces the piecewise linear growth
functions which are non-smooth and very far from being linear. This
nonlinearity transforms normal or uniform distributions of resource
availabilities into skewed crop yield distribution and no natural
satisfactory motivation exists in favor of any simple crop yield
distribution \cite{Hennessy2009}. With independent, identical,
uniform resource availability distributions the yield skew is
positive, and it is negative for normal distributions.

The standard linear tools of statistics such as generalized linear
models do not work satisfactory for systems with limiting factors.
Conventional correlation analysis conflicts with the concept of
limiting factors. This was demonstrated in a study of the spatial
distribution of Glacier lily in relation to soil properties and
gopher disturbance \cite{Thomsonatal1996}. For systems with limiting
factors, quantile regression performs much better with strong
theoretical justification in Law of the Minimum \cite{Austin2007}.

Some of the generalizations of the Law of the Minimum went quite far
from agriculture and ecology. The Law of the Minimum was applied to
economics \cite{EcolEcon} and to education, for example
\cite{EcolEdu}.

Recently, a strong mathematical background was created for the Law
of the Minimum. Now the limiting factors theory together with static
and dynamic limitation in chemical kinetics
\cite{GorRadLim,GorbanRadZin2010} are considered as the realization
of the {\it Maslov dequantization}
\cite{KoMa97,LitvinovMaslov2005,Litvinov2007} and {\it idempotent
analysis}. Roughly speaking, the limiting factor formalism means
that we should handle any two quantities $c_1,c_2$ either as equal
numbers or as numbers connected by the relation $\gg$: either
$c_1\gg c_2$ or $c_1 \ll c_2$. Such a hard non-linearity can arise
in the smooth dynamic models because of the time-scale separation
\cite{vandenBerg1998}.

Dequantization of the traditional mathematics leads to a mathematics
over tropical algebras like the max-plus algebra.  Since the
classical work of Kleene \cite{Kleene} these algebras are
intensively used in mathematics and computer science, and the
concept of dequantization and idempotent analysis opened new
applications in physics and other natural sciences (see the
comprehensive introduction in \cite{Litvinov2007}). Liebig's and
anti-Liebig's (see Definition 1 below) systems of factors may be
considered as realizations of max-plus or min-plus asymptotics
correspondingly.

\subsection{Fitness Convexity, Concavity and Various Interactions Between
Factors}

There exist an opposite type of organization of the system of
factors, which, from a first glance, seems to be symmetric to
Liebig's type of interaction between them. In Liebig's systems, the
factor with the worst value determines the growth and surviving. The
completely opposite situation is: the factor with the best value
determines everything. We call such a system ``anti-Liebig's" one.
Of course, it seems improbable that all the possible factors
interact following the Law of the Minimum or the fully opposite
anti-Liebig's rule. Interactions between factors in real systems are
much more complicated \cite{SaitoGoepfert2008}. Nevertheless, we can
state a question about hierarchical decomposition of the system of
factors in elementary groups with simple interactions inside, then
these elementary groups can be clustered into super--factors with
simple interactions between them, and so on.

Let us introduce some notions and notations. We consider organisms
that are under the influence of several factors $F_1,... F_q$. Each
factor has its intensity $f_i$ ($i=1,...q$). For convenience, we
consider all these factors as negative or  harmful. This is just a
convention about the choice of axes directions: a wholesome factor
is just a ``minus harmful" factor.

At this stage, we do not specify the nature of these factors.
Formally, they are just inputs in the adaptation dynamics, the
arguments of the {\em fitness functions}.

The fitness function is the central notion of the evolutionary and
ecological dynamics. This is a function that maps the environmental
factors and traits of the organism into the reproduction
coefficient, that is, its contribution, in offspring to its
population. Fisher proposed to construct fitness as a combination of
independent individual contribution of various traits
\cite{Fisher1930}. Haldane \cite{Haldane1932} criticized the
approach based on independent actions of traits. Modern definitions
of fitness function are based on adaptation dynamics. For the
structured populations, the fitness should be defined through the
dominant Lyapunov exponents \cite{G1984,MetzNisbetGeritz1992}. In
the evolutionary game theory \cite{Maynard-Smith1982}, payoff
represents Darwinian fitness and describes how the use of the
strategy improves an animal's prospects for survival and
reproduction. Recently, the Fisher and Haldane approaches are
combined \cite{WaxmanWelch2005}: Haldane's concern is incorporated
into Fisher's model by allowing the intensity of selection to vary
between traits.

It is a  nontrivial task to measure the fitness functions and action
of selection in nature, but now it has been done for many
populations and phenotypical traits \cite{KingsolverPfennig2007}.
Special statistical methods for life-history analysis for inference
of fitness and population growth are developed and tested
\cite{Shawatal2008}.

In our further analysis we do not need exact values of fitness but
rather its existence and some qualitative features.

First of all, let us consider an oversimplified situation with
identical organisms. Given phenotypical treats, fitness $W$ is a
function of factor loads: $W=W(f_1, \ldots, f_q)$. This assumption
does not take into account physiological adaptation that works as a
protection system and modifies the factor loads. This modification
is in the focus of our analysis in the follow-up section, but for
now we neglect adaptation. The convention about axes direction means
that all the partial derivatives of $W$ are non-positive $\partial W
/\partial f_i \leq 0$.

\begin{figure}
\centering{ \subfloat[Liebig's
system]{\label{Liebig-}\includegraphics[width=0.4\textwidth]{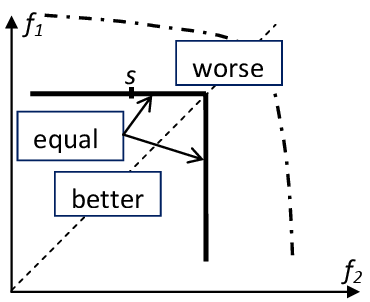}}
\quad \subfloat[Generalized Lie\-big's sys\-tem] {\label{GenLiebig}
\includegraphics[width=0.4\textwidth]{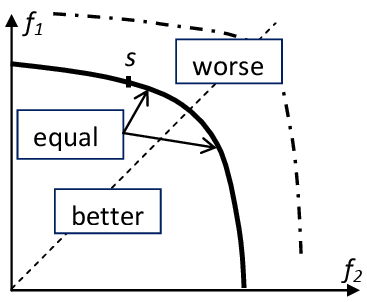}} \\
\subfloat[Anti-Lie\-big's system] {\label{AntiLiebig}
\includegraphics[width=0.4\textwidth]{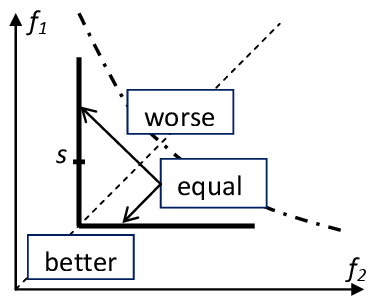}}\quad
\subfloat[Synergistic system] {\label{Synerg}
\includegraphics[width=0.4\textwidth]{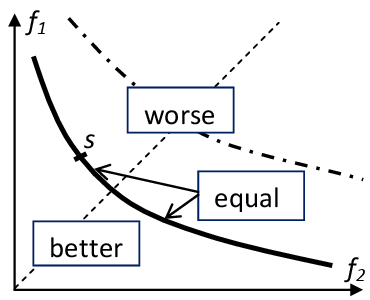} }}\caption{Various
types of organization of the system of factors. For a given state
$s$ the bold solid line is given by the equation $W(f_1,f_2)=W(s)$.
This line separates the area with higher fitness (``better
conditions") from the line with lower fitness (``worse conditions").
In Liebig's (a) and generalized Liebig's systems (b) the area of
better conditions is convex, in ``anti-Liebig's" systems (c) and the
general synergistic systems (d)  the area of worse conditions is
convex. The dot dash line shows the border of survival. On the
dashed line the factors are equally important ($f_1=f_2$).
\label{VarOgan}}
\end{figure}

By definition, for a {\it Liebig's system} of factors $W$ is a
function of the worst (maximal) factor intensity: $W=W(\max\{f_1,
\ldots , f_q\})$  (Fig.~\ref{Liebig-}) and for {\it anti-Liebig's}
system it is the function of the best (minimal) factor intensity
$W=W(\min\{f_1, \ldots , f_q\})$ (Fig.~\ref{AntiLiebig}). Such
representations as well as the usual formulation of the Law of the
Minimum require special normalization of factor intensities to
compare the loads of different factors.

For Liebig's systems of factors the superlevel sets of $W$
given by inequalities $W \geq w_0$ are convex for any level
$w_0$ in a convex domain (Fig.~\ref{Liebig-}). For
anti-Liebig's systems of factors the sublevel sets of $W$ given
by inequalities $W \leq w_0$ are convex for any level $w_0$ in
a convex domain (Fig.~\ref{AntiLiebig}).

These convexity properties are essential for optimization problems
which arise in the modeling of adaptation and evolution. Let us take
them as definitions of the generalized Liebig and anti-Liebig
systems of factors:

\begin{definition} \label{Def1}
\begin{enumerate}
\item{A system of factors is the {\em generalized Liebig
    system} in a convex domain $U$, if for any level $w_0$
    the superlevel set $\{f \in U \ | \ W(f) \geq w_0 \}$
    is convex (Fig.~\ref{GenLiebig}).}
\item{A system of factors is the {\em generalized anti-Liebig
    system} in a convex domain $U$, if for any level $w_0$
    the sublevel set $\{f \in U \ | \ W(f) \leq w_0 \}$ is
    convex (Fig.~\ref{Synerg})}.
\end{enumerate}
\end{definition}
We call the generalized anti-Liebig systems of factors the {\em
synergistic} systems because this formalizes the idea of synergy: in
the synergistic systems harmful factors superlinear amplify each
other.

\begin{figure}
\subfloat[Liebig's
system]{\label{Liebig-Optimal}\includegraphics[height=0.3\textwidth]{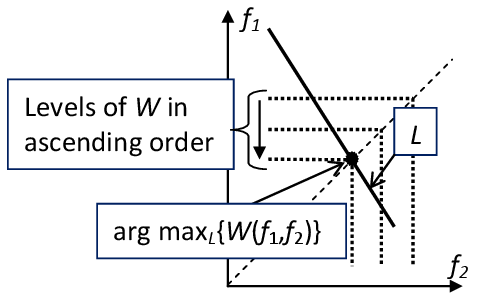}}
\quad \subfloat[Generalized Lie\-big's sys\-tem]
{\label{GenLiebigOptimal}
\includegraphics[height=0.3\textwidth]{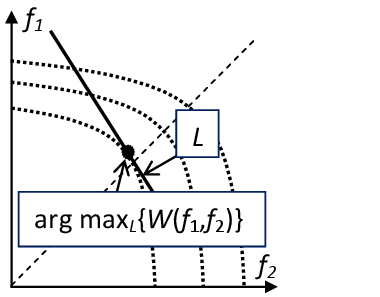}} \\
\subfloat[Anti-Lie\-big's system] {\label{AntiLiebigOptimal}
\includegraphics[height=0.3\textwidth]{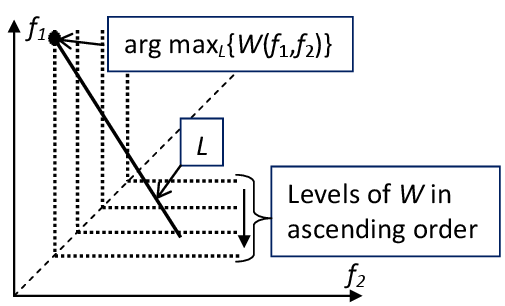}}\quad
\subfloat[Synergistic system] {\label{SynergOPtimal}
\includegraphics[height=0.3\textwidth]{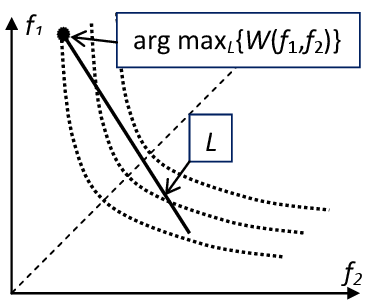} }
\caption{Conditional optimization for various systems of factors.
Because of convexity conditions, fitness achieves its maximum on an
interval $L$ for Liebig's system (a) on the diagonal (the factors
are equally important), for generalized Liebig's systems (b) near
the diagonal, for anti-Liebig's system (c) and for the general
synergistic system (d) this maximum is one of the ends of the
interval $L$.  \label{Optimizm}}
\end{figure}

Conditional maximization of fitness destroys the symmetry between
Liebig's and anti-Liebig's systems as well as between generalized
Liebig's systems and synergistic ones. Following the geometric
approach of \cite{Tilman1980,Tilman1982} we illustrate this
optimization on Fig.~\ref{Optimizm}. The picture may be quite
different from the conditional maximization of a convex function
near its minima point (compare, for example,
Figs.~\ref{AntiLiebigOptimal},~\ref{SynergOPtimal} to Fig. from
\cite{SihGleeson1995}).

Individual adaptation changes the picture. In the next subsection we
discuss possible mechanism of these changes.

\subsection{Adaptation Energy and Factor--Resource Models}

The reaction of an organism to the load of a single factor may have
plateaus (intervals of tolerance considered in Shelford's ``law of
tolerance", \cite{Odum}, Chapter 5). The dose--response curves may
be nonmonotonic \cite{Colborn} or even oscillating. Nevertheless, we
start from a very simple abstract model that is close to the usual
factor analysis.

We consider organisms that are under the influence of several
harmful factors $F_1,... F_q$ with  intensities $f_i$ ($i=1,...q$).
Each organism has its adaptation systems, a ``shield" that can
decrease the influence of external factors. In the simplest case, it
means that each system has an available adaptation resource, $R$,
which can be distributed for the neutralization of factors: instead
of factor intensities $f_i$ the system is under pressure from factor
values $f_i-a_i r_i$ (where $a_i>0$ is the coefficient of efficiency
of factor $F_i$ neutralization by the adaptation system and $r_i$ is
the share of the adaptation resource assigned for the neutralization
of factor $F_i$, $\sum_i r_i \leq R$). The zero value $f_i-a_i r_i=
0$ is optimal (the fully compensated factor), and further
compensation is impossible and senseless.

For unambiguity of terminology, we use the term ``factor" for all
factors including any deficit of available external resource or even
some illnesses. We keep the term ``resource" for internal resources,
mostly for the hypothetical Selye's ``adaptation energy".

It should be specially stressed that the adaptation energy is
neither physical energy nor a substance. This idealization describes
the experimental results: in many experiments it was demonstrated,
that organisms under load of various factors {\em behave as} if they
spend a resource, which is the same for different factors. This
resource may be exhausted and then the organism dies.

We represent the organisms, which are adapting to stress, as the
systems which optimize distribution of available amount of a special
adaptation resource for neutralization of different aggressive
factors (we consider the deficit of anything necessary as a negative
factor too). These {\it factor--resource} models with optimization
are very convenient for the modeling of adaptation. We use a class
of models {\it many factors -- one resource}.

Interaction of  each system with a factor $F_i$ is described by
two quantities: the factor $F_i$ pressure $\psi_i=f_i-a_i r_i$ and
the resource $r_i$ assigned to the factor $F_i$ neutralization.
The first quantity characterizes, how big the uncompensated harm
is from that factor, the second quantity measures, how intensive
is the adaptation answer to the factor (or how far the system was
modified to answer the factor $F_i$ pressure).

Already one factor--one resource models of adaptation produce the
tolerance law. We demonstrate below that it predicts the separation
of groups of organism into two subgroups: the less correlated
well--adapted organisms and highly correlated organisms with a
deficit of the adaptation resource. The variance is also higher in
the highly correlated group of organisms with a deficit of the
adaptation resource.

This result has a clear geometric interpretation. Let us represent
each organism as a data point in an $n$-dimensional vector space.
Assume that they fall roughly within an ellipsoid. The well-adapted
organisms are not highly correlated and after normalization of
scales to unit variance the corresponding cloud of points looks
roughly as a sphere. The organisms with a deficit of the adaptation
resource are highly correlated, hence in the same coordinates their
cloud looks like an ellipsoid with remarkable eccentricity.
Moreover, the largest diameter of this ellipsoid is larger than for
the well--adapted organisms and the variance increases together with
the correlations.

This increase of variance together with correlations may seem
counterintuitive because it has no formal backgrounds in definitions
of the correlation coefficients and variance. This is an empirical
finding that under stress correlations and variance increase
together, supported by many observations both for physiological and
financial systems. The factor-resource models give a plausible
explanation of this phenomena.

The crucial question is: what is the {\it resource of adaptation}?
This question arose for the first time when Selye published the
concept of {\it adaptation energy} and experimental evidence
supporting this idea \cite{SelyeAEN,SelyeAE1}. Selye found that the
organisms (rats) which demonstrate no differences in normal
environment may differ significantly in adaptation to an increasing
load of environmental factors. Moreover, when he repeated the
experiments, he found that adaptation ability decreases after
stress. All the observations could be explained by existence of an
universal adaptation resource that is being spent during all
adaptation processes.

Selye's ideas allow the following interpretation: the aggressive
influence of the environment on the organism may be represented as
an action of independent factors. The system of adaptation consists
of subsystems, which protect the organism from different factors.
These subsystems consume the same resource, the adaptation energy.
The distribution of this resource between the subsystems depends on
environmental conditions.

Later the concept of adaptation energy was significantly improved
\cite{GP_AE1952}, plenty of indirect evidence supporting this
concept were found, but this elusive adaptation energy is still a
theoretical concept, and in the modern ``Encyclopedia of Stress" we
read: ``As for adaptation energy, Selye was never able to measure
it..." \cite{AEencicl}. Nevertheless, the notion of adaptation
energy is very useful in the analysis of adaptation and is now in
wide use (see, for example,
\cite{BreznitzAEappl,SchkadeOccAdAE2003}).

The idea of {\it exchange} can help in the understanding of
adaptation energy: there are many resources, but any resource can be
exchanged for another one. To study such an exchange an analogy with
the currency exchange is useful. Following this analogy, we have to
specify, what is the {\it exchange rate}, how fast this exchange
could be done (what is the {\it exchange time}), what is the {\it
margin}, and how the margin depends on the exchange time. There may
appear various limitations of the amount of the exchangeable
resource, and so on. The economic metaphor for ecological
concurrency and adaptation was elaborated in 1985
\cite{BloomChapinMooney1985,ChapinSchulzeMooney1990} but much
earlier, in 1952, it was developed for physiological adaptation
\cite{GP_AE1952}.

Market economics seems closer to the idea of resource
universalization than biology is, but for biology this exchange idea
also seems useful. Of course there exist some limits on the possible
exchanges of different resources. It is possible to include the
exchange processes into models, but many questions arise about
unknown coefficients. Nevertheless, we can follow Selye's arguments
and postulate the adaptation energy as a universal adaptation
resource.

The adaptation energy is neither physical energy nor a substance.
This is a theoretical construction, which may be considered as a
pool of various exchangeable resources. When an organism achieves
the limits of resource exchangeability, the universal non-specific
stress and adaptation syndrome transforms (disintegrates) into
specific diseases. Near this limit we have to expect the {\it
critical retardation} \cite{GorSlowRelax} of exchange processes.

Adaptation optimizes the state of the system for given available
amounts of the adaptation resource. This idea seems very natural,
but it may be a difficult task to find the objective function that
is hidden behind the adaptation process. Nevertheless, even an
assumption about the existence of an objective function and about
its general properties helps in analysis of adaptation process.

Assume that adaptation should maximize a fitness function $W$
which depends on the compensated values of factors,
$\psi_i=f_i-a_i r_i$ for the given amount of available
resource:
\begin{equation}\label{Optimality}
\left\{ \begin{array}{l}
 W(f_1-a_1 r_1, f_2-a_2 r_2, ... f_q-a_q r_q) \ \to \ \max \ ; \\
 r_i \geq 0$, $f_i-a_i r_i \geq 0$, $\sum_{i=1}^q r_i \leq R \ .
\end{array} \right.
\end{equation}

The only question is: how can we be sure that adaptation follows any
optimality principle? Existence of optimality is proven for
microevolution processes and ecological succession. The mathematical
backgrounds for the notion of ``natural selection" in these
situations are well--established after work by Haldane (1932)
\cite{Haldane1932} and Gause (1934) \cite{Gause}. Now this direction
with various concepts of {\it fitness} (or ``generalized fitness")
optimization is elaborated in many details (see, for example, review
papers \cite{Bom02,Oechssler02,GorbanSelTth}).

The foundation of optimization is not so clear for such processes as
modifications of a phenotype, and for adaptation in various time
scales. The idea of {\it genocopy--phenocopy interchangeability} was
formulated long ago by biologists to explain many experimental
effects: the phenotype modifications simulate the optimal genotype
(\cite{West-Eberhardgenocopy-phenocopy}, p. 117). The idea of
convergence of genetic and environmental effects was supported by an
analysis of genome regulation \cite{ZuckerkandlConvergGenEnv} (the
principle of concentration--affinity equivalence). The phenotype
modifications produce the  same change, as evolution of the genotype
does, but faster and in a smaller range of conditions (the proper
evolution can go further, but slower). It is natural to assume that
adaptation in different time scales also follows the same direction,
as evolution and phenotype modifications, but faster and for smaller
changes. This hypothesis could be supported by many biological data
and plausible reasoning. (See, for example, the case studies of
relation between evolution of physiological adaptation
\cite{Hoffman1978,Greene1999}, a book about various mechanisms of
plants responses to environmental stresses \cite{Lerner1999}, a
precise quantitative study of the relationship between evolutionary
and physiological variation in hemoglobin \cite{Miloatal2007} and a
modern review with case studies \cite{FuscoMinelli2010}.)

It may be a difficult task to find an explicit form of the fitness
function $W$, but for our qualitative analysis we need only a
qualitative assumption about general properties of $W$. First, we
assume monotonicity with respect to each coordinate:
\begin{equation}\label{monotonicity}
\frac{\partial W(\psi_1, \ldots \psi_q)}{\partial \psi_i} \leq 0
\, .
\end{equation}
A system of factors is {\it Liebig's system}, if
\begin{equation}\label{objective}
W=W\left(\max_{1\leq i\leq q} \{f_i-a_i r_i\}\right)\ .
\end{equation}
This means that fitness depends on the worst factor pressure.

A system of factors is generalized Liebig's system (Definition 1.1),
if for any two different vectors of factor pressures
$\mathbf{\psi}=(\psi_1,... \psi_q)$ and $\mathbf{\phi}=(\phi_1,...
\phi_q)$ ($\mathbf{\psi} \neq \mathbf{\phi}$)  the value of fitness
at the average point $(\mathbf{\psi}+\mathbf{\phi})/2$ is greater,
than at the worst of points $\mathbf{\psi}$, $\mathbf{\phi}$:
\begin{equation}\label{GenLiebigEQ}
W\left(\frac{\mathbf{\psi}+\mathbf{\phi}}{2}\right) > \min
\{W(\mathbf{\psi}),W(\mathbf{\phi})\}\ .
\end{equation}

Any Liebig's system is, at the same time, generalized Liebig's
system because for such a system the fitness  $W$ is a decreasing
function of the maximal factor pressure, the minimum of $W$
corresponds to the maximal value of the limiting factor and
$$\max\left\{\frac{\psi_1+\phi_1}{2}, \ldots
,\frac{\psi_q+\phi_q}{2}\right\} \leq \max \{ \max \{\psi_1, \ldots
, \psi_q\}, \max \{\phi_1, \ldots , \phi_q\}\} \ .$$

The opposite principle of factor organization is synergy: the
superlinear mutual amplification of factors. The system of factors
is a {\it synergistic} one (Definition 1.2), if for any two
different vectors of factor pressures $\mathbf{\psi}=(\psi_1,...
\psi_q)$ and $\mathbf{\phi}=(\phi_1,... \phi_q)$ ($\mathbf{\psi}
\neq \mathbf{\phi}$)  the value of fitness at the average point
$(\mathbf{\psi}+\mathbf{\phi})/2$ is less, than at the best of
points $\mathbf{\psi}$, $\mathbf{\phi}$:
\begin{equation}\label{synergy}
W\left(\frac{\mathbf{\psi}+\mathbf{\phi}}{2}\right) <
\max\{W(\mathbf{\psi}),W(\mathbf{\phi})\}\ .
\end{equation}
A system of factors is {\it anti-Liebig's system}, if
\begin{equation}\label{Anty-objective}
W=W\left(\min_{1\leq i\leq q} \{f_i-a_i r_i\}\right)\ .
\end{equation}
This means that fitness depends on the best factor pressure. Any
anti-Liebig system is, at the same time a synergistic one because
for such a system the fitness $W$ is a decreasing function of the
minimal factor pressure, the maximum of $W$ corresponds to the
minimal value of the factor with minimal pressure and

$$\min\left\{\frac{\psi_1+\phi_1}{2}, \ldots
,\frac{\psi_q+\phi_q}{2}\right\} \geq \min \{ \min \{\psi_1, \ldots
, \psi_q\}, \min \{\phi_1, \ldots , \phi_q\}\}$$

We prove that adaptation of an organism to Liebig's system of
factors, or to any synergistic system, leads to two paradoxes of
adaptation:

\begin{itemize}
\item{{\it Law of the Minimum paradox} (Sec.~ref{Sec:LawMinParad}):
If for a randomly selected pair, ("State of environment -- State of
organism"), the Law of the Minimum is valid (everything is limited
by the factor with the worst value) then, after adaptation, many
factors (the maximally possible amount of them) are equally
important. }
\item{{\it Law of the Minimum inverse paradox} (Sec.~ref{Sec:LawMinInvPar}): If for a randomly
selected pair, ("State of environment -- State of organism"), many
factors are equally important  and superlinearly amplify each other
then, after adaptation, a smaller amount of factors is important
(everything is limited by the factors with the worst non-compensated
values, the system approaches the Law of the Minimum).}
\end{itemize}
 In this paper, we discuss the individual adaptation. Other types
of adaptations, such as changes of the ecosystem structure,
ecological succession or microevolution lead to the same paradoxes
if the factor--resource models are applicable to these processes.

\section{One-Factor Models, the Law of Tolerance, and the Order--Disorder Transition}

The question about interaction of various factors is very
important, but, first of all, let us study the one-factor models.
Each organism is characterized by measurable attributes $x_1,
\ldots x_m$ and the value of adaptation resource, $R$.

\subsection{Tension--Driven Models}

In these models, observable properties of interest $x_k$
$(k=1,...m)$ can be modeled as functions of the pressure factor
$\psi$ plus some noise $\epsilon_k$.

Let us consider one-factor systems and linear functions (the
simplest case). For the tension--driven model the attributes $x_k$
are linear functions of tension $\psi$ plus noise:
\begin{equation}\label{1factorTension}
x_k=\mu_k + l_k \psi+ \epsilon_k \ ,
\end{equation}
where $\mu_k$ is the expectation of $x_k$ for fully compensated
factor, $l_k$ is a coefficient, $\psi=f-ar_f \geq 0$, and $r_f
\leq R$ is amount of available resource assigned for the factor
neutralization. The values of $\mu_k$ could be considered as
``normal" (in the sense opposite to ``pathology"), and noise
$\epsilon_k$ reflects variability of norm.

If systems compensate as much of factor value, as it is possible,
then $r_f= \min\{R,f/a\}$, and we can write:
\begin{equation}\label{OneFactor}
 \psi=\left\{\begin{array}{ll}
 &f-a R\ , \ \ {\rm if} \ \ f>aR \ ; \\
 &0 ,\ \  \ {\rm else.}
 \end{array}
 \right.
\end{equation}

Individual systems may be different by the value of factor intensity
(the local intensity variability), by amount of available resource
$R$ and, of course, by the random values of $\epsilon_k$. If all
systems have enough resource for the factor neutralization ($aR>f$)
then all the difference between them is in the noise variables
$\epsilon_k$. No change will observed under increase of the factor
intensity, until violation of inequality $F<r$ occurs.

Let us define the dose--response curve as
$$M_k(f)=\mathbf{E}(x_k|f).$$ Due to (\ref{1factorTension})
\begin{equation}\label{1factorDose-Rsp}
M_k(f)=\mu_k + l_k \mathbf{P}(aR<f)(f-a\mathbf{E}(R|aR<f)) \ ,
\end{equation}
where $\mathbf{P}(aR<f)$ is the probability of organism to have
insufficient amount of resource for neutralization of the factor
load and $\mathbf{E}(R|aR<f)$ is the conditional expectation of
the amount of resource if it is insufficient.

The slope $\D M_k(f)/ \D f $ of the dose--response curve
(\ref{1factorDose-Rsp}) for big values of $f$ tends to $l_k$, and
for small $f$ it could be much smaller. This plateau at the
beginning of the dose-response curve corresponds to the law of
tolerance  (V.E. Shelford, 1913, \cite{Odum}, Chapter 5).

If the factor value increases, and for some of the systems the
factor intensity $f$ exceeds the available compensation $a R$ then
for these systems $\psi>0$ and the term $l_k \psi$ in Eq.
(\ref{1factorTension}) becomes important. If the noise of the norm
$\epsilon_k$ is independent of $\psi$ then the correlation between
different $x_k$ increases monotonically with $f$.

With increase of the factor intensity $f$ the dominant eigenvector
of the correlation matrix between $x_k$ becomes more uniform in
the coordinates, which tend asymptotically to $\pm
\frac{1}{\sqrt{m}}$.

For a given value of the factor intensity $f$ there are two groups
of organisms: the well--adapted group with $R\geq f$ and $\psi=0$,
and the group of organisms with deficit of adaptation energy and
$\psi>0$. If the fluctuations of norm $\epsilon_k$ are independent
for different $k$ (or just have small correlation coefficients),
then in the group with deficit of adaptation energy the correlation
between attributes is much higher than in the well--adapted group.
If we use a metaphor from physics, we can call these two groups {\it
two phases}: the highly correlated phase with deficit of adaptation
energy and the less correlated phase of well-adapted organisms.

In this simple model (\ref{1factorTension}) we just formalize
Selye's observations and theoretical argumentation. One can call
it {\it Selye's model}. There are two other clear possibilities
for one factor--one resource models.

\subsection{Response--Driven Models}

What is more important for values of the observable quantities
$x_k$: the current pressure of the factors, or the adaptation to
this factor which modified some of parameters? Perhaps, both, but
let us introduce now the second simplest model.

In the response--driven model of adaptation, the quantities $x_k$
are modeled as linear functions of adaptive response $ar_f$ (with
coefficients $q_k$) plus some noise $\epsilon_k$:
\begin{equation}\label{1factorModification}
x_k=\mu_k + q_k a r_f + \epsilon_k \ .
\end{equation}
When $f$ increases then, after threshold $f=a R$, the term $q_k a
r_f$ transforms into $q_k a R$ and does not change further. The
observable quantities $x_k$ are not sensitive to changes in the
factor intensity $f$ when $f$ is sufficiently large. This is the
significant difference from the behavior of the tension-driven model
(\ref{1factorTension}), which is not sensitive to change of $f$ when
$f$ is sufficiently small.

\subsection{Tension--and--Response Driven 2D One--Factor Models}

This model is just a linear combination of
Eqs.~(\ref{1factorTension}) and (\ref{1factorModification})
\begin{equation}\label{1factorTen+Mod}
x_k=\mu_k + l_k \psi + q_k a r_f + \epsilon_k \ .
\end{equation}
For small $f$ (comfort zone) $\psi=0$, the term $l_k \psi$
vanishes, $a r_f=f$ and the model has the form $x_k=\mu_k + q_k f
+ \epsilon_k $. For intermediate level of $f$, if systems with
both signs of inequality $f \gtreqless a R$ are present, the model
imitates 2D (two-factor) behavior. After the threshold $f \geq a
R$ is passed for all systems, the model demonstrates 1D behavior
again: $x_k=\mu_k + l_k f + (q_k-l_k) a R + \epsilon_k \ .$ For
small $f$ the motion under change of $f$ goes along direction
$q_k$, for large $f$ it goes along direction $l_k$.

Already the first model of adaptation (\ref{1factorTension}) gives
us the law of tolerance and practically important effect of
order--disorder transition under stress. Now we have no arguments
for decision which of these models is better, but the second model
(\ref{1factorModification}) has no tolerance plateau for small
factor values, and the third model has almost two times more
fitting parameters. Perhaps, the first choice should be the first
model (\ref{1factorTension}), with generalization to
(\ref{1factorTen+Mod}), if the described two-dimensional behaviour
is observed.

\section{Law of the Minimum Paradox \label{Sec:LawMinParad}}

Liebig used the image of a barrel -- now called Liebig's barrel --
to explain his law. Just as the capacity of a barrel with staves
of unequal length is limited by the shortest stave, so a plant's
growth is limited by the nutrient in shortest supply.

Adaptation system acts as a cooper and repairs the shortest stave to
improve the barrel capacity. Indeed, in well-adapted systems the
limiting factor should be compensated as far as this is possible. It
seems obvious because of the very natural idea of optimality, but
arguments of this type in biology should be considered with care.

Assume that adaptation should maximize a objective function $W$
(\ref{Optimality}), which satisfies the Law of the Minimum
(\ref{objective}) and the monotonicity requirement
(\ref{monotonicity}) under conditions $r_i \geq 0$, $f_i-a_i r_i
\geq 0$, $\sum_{i=1}^q r_i \leq R$. (Let us remind that $f_i \geq 0$
for all $i$.)

Description of the maximizers of $W$ gives the following theorem.

\begin{theorem} \label{Theorem1} For any objective function $W$ that satisfies conditions
(\ref{objective}) the optimizers $r_i$ are defined by the following
algorithm.
\begin{enumerate}
\item{Order intensities of factors: $f_{i_1} \geq f_{i_1} \geq ...
f_{i_q}$.}
 \item{Calculate differences $\Delta_j =f_{i_j} -f_{i_{j+1}}$
(take formally $\Delta_0=\Delta_{q+1}=0$).}
 \item{Find such $k$ ($0 \leq
k \leq q$) that
  $$\sum_{j=1}^{k} \left(\sum_{p=1}^j \frac{1}{a_{i_p}}\right) \Delta_j
  \leq R \leq \sum_{j=1}^{k+1} \left(\sum_{p=1}^j
 \frac{1}{a_{i_p}}\right) \Delta_j \ . $$ For $R< \Delta_1$
 we put $k=0$ and if  $R>  \sum_{j=1}^q \left(\sum_{p=1}^j
 \frac{1}{a_{i_p}}\right) \Delta_j$ then we take $k=q$.}
\item{If $k < q$ then the optimal amount of resource $r_{i_j}$ is: for $j=1, \ldots , k+1$
\begin{equation}\label{OptimDistr}
r_{i_j}=\frac{f_{i_j}-\psi}{a_{i_j}}\, , \ {\rm where} \
\psi=\left(\sum_{p=1}^{k+1}\frac{1}{a_{i_p}}\right)^{-1}
\left(\sum_{p=1}^{k+1}\frac{f_{i_p}}{a_{i_p}}-R\right)\,
\end{equation}
and $r_{i_j}=0$ for $j>k+1$. If $k=q$ then $r_i = f_i/a_i$ for all
$i$.}
\end{enumerate}
\end{theorem}

\begin{figure} \centering{
\includegraphics[width=100mm]{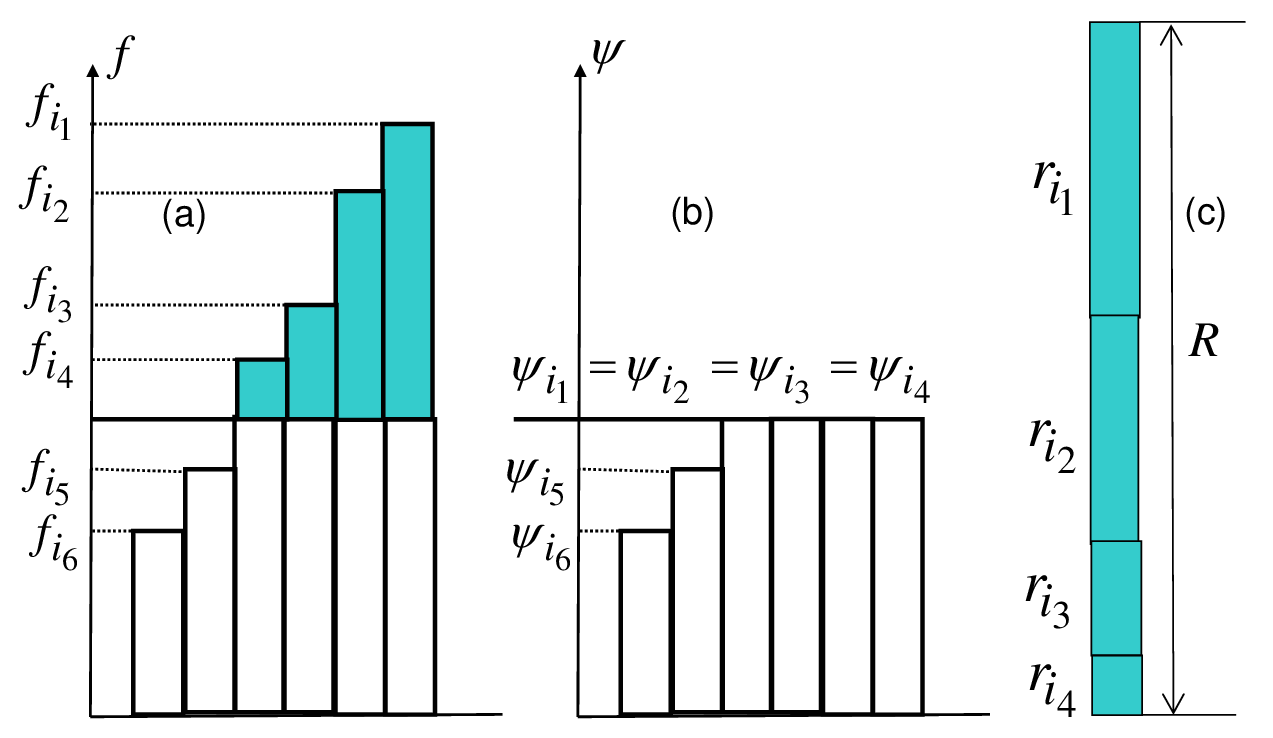}}
\caption{\label{Fig:FactorDistrib} Optimal distribution of resource
for neutralization of factors under the Law of the Minimum. (a)
histogram of factors intensity (the compensated parts of factors are
highlighted, $k=3$), (b) distribution of tensions $\psi_i$ after
adaptation becomes more uniform, (c) the sum of distributed
resources. For simplicity of the picture, we take here all $a_i=1$.}
\end{figure}

\begin{proof} This optimization is illustrated in
Fig.~\ref{Fig:FactorDistrib}. If $R \geq \sum_i f_{i_j}/a_{i_j}$
then the pressure of all the factors could be compensated and we can
take $r_i=f_i/a_i$. Now, let us assume that $R < \sum_i
f_{i_j}/a_{i_j}$. In this case, the pressure of some of the factors
is not fully compensated. The adaptation resource is spent for
partial compensation of the $k+1$ worst factors and the remained
pressure of them is higher (or equal) then the pressure of the
($k+2$)nd worst factor $F_{i_{k+2}}$:
\begin{equation}\label{extrcond}
\begin{split}
&f_{i_1}-a_{i_1}r_{i_1}=\ldots
=f_{i_{k+1}}-a_{i_{k+1}}r_{i_{k+1}}=\psi \geq
f_{i_{k+2}}\, , \; \sum_{j=1}^{k+1} r_{i_j}=R \, , \ \mbox{and} \\
&\sum_{i=1}^{k+1} \Delta_i -a_{i_1}r_{i_1} = \ldots = \Delta_{k+1} -
a_{i_{k+1}}r_{i_{k+1}} = \psi-f_{i_{k+2}}=\theta_{k+1} \geq 0 \, .
\end{split}
\end{equation}
Therefore, for $j=1,\ldots , k+1$ in the optimal distribution of the
resource,
\begin{equation}
r_{i_j}= \frac{1}{a_{i_j}}\left(\sum_{i=j}^{k+1}\Delta_i -
\theta_{k+1}\right)\, , \ R=\sum_{j=1}^{k+1}r_{i_j}\, , \
 \theta_{k+1}\geq 0\, .
\end{equation}
This gives us the first step in the Theorem~\ref{Theorem1}, the
definition of $k$. Formula (\ref{OptimDistr}) for $r_{i_j}$ follows
also from (\ref{extrcond}).
\end{proof}

Hence, if the system satisfies the Law of the Minimum then the
adaptation process makes the tension produced by different factors
more uniform (Fig.~\ref{Fig:FactorDistrib}). Thus adaptation
decreases the effect from the limiting factor and hides
manifestations of the Law of the Minimum.

Under the assumption of optimality  (\ref{Optimality}) the {\it Law
of the Minimum paradox} becomes a theorem: if the Law of the Minimum
is true then microevolution, ecological succession, phenotype
modifications and adaptation decrease the role of the limiting
factors and bring the tension produced by different factors
together.

The cooper starts to repair Liebig's barrel from the shortest stave
and after reparation the staves are more uniform, than they were
before. This cooper may be microevolution, ecological succession,
phenotype modifications, or adaptation. For the ecological
succession this effect (the Law of the Minimum leads to its
violation by succession) was described in Ref.~\cite{SemSem}. For
adaptation (and in general settings too) it was demonstrated in
Ref.~\cite{GorSmiCorAd1st}.

\section{Law of the Minimum Inverse Paradox
\label{Sec:LawMinInvPar}}

The simplest formal example of ``anti--Liebig's" organization of
interaction between factors gives us the following dependence of
fitness from two factors: $W=-f_1 f_2$: each of factors is neutral
in the absence of another factor, but together they are harmful.
This is an example of {\it synergy}: the whole is greater than the
sum of its parts. (For our selection of axes direction, ``greater"
means ``more harm".)

In according to Definition~\ref{Def1}, the system of factors
$F_1,... F_q$ is synergistic, in a convex domain $U$ of the
admissible vectors of factor pressure if for any level $w_0$ the
sublevel set $\{\psi \in U \ | \ W(\psi) \leq w_0 \}$ is convex.
Another definition gives us the synergy inequality (\ref{synergy}).
These definitions are equivalent. This proposition follows from the
definition of convexity and standard facts about convex sets (see,
for example, \cite{Rockafellar})

\begin{proposition} \label{Prop1} The synergy inequality (\ref{synergy})
holds if and only if all the sublevel sets $\{\mathbf{f} \ | \
W(\mathbf{f}) \leq \alpha \}$ are strictly convex.
\end{proposition}
(The fitness itself may be a non-convex function.)

This proposition immediately implies that the synergy inequality
is invariant with respect to increasing monotonic transformations
of $W$. This invariance with respect to nonlinear change of scale
is very important, because usually we don't know the values of
function $W$.

\begin{proposition} \label{Prop2}If the synergy inequality (\ref{synergy})
holds for a function $W$, then it holds for a function
$W_{\theta}=\theta (W)$, where $\theta(x)$ is an arbitrary strictly
monotonic function of one variable.
\end{proposition}

Already this property allows us to study the problem about optimal
distribution of the adaptation resource without further knowledge
about the fitness function.

Assume that adaptation should maximize an objective function
$W(f_1-r_1, ... f_q-r_q)$  (\ref{Optimality}) which satisfies the
synergy inequality (\ref{synergy}) under conditions $r_i \geq 0$,
$f_i-a_i r_i \geq 0$, $\sum_{i=1}^q r_i \leq R$. (Let us remind that
$f_i \geq 0$ for all $i$.) Following our previous convention about
axes directions all factors are harmful and $W$ is monotonically
decreasing function (\ref{monotonicity}). We need also a technical
assumption that $W$ is defined on a convex set in $\mathbb{R}^q_+$
and if it is defined for a nonnegative point $\mathbf{f}$, then it
is also defined at any nonnegative point $\mathbf{g} \leq
\mathbf{f}$ (this inequality means that $g_i \leq f_i$  for all
$i=1,... q$).

The set of possible maximizers is finite. For every group of $j+1$
factors ($1 \leq j+1 < q$), $F_{i_1},... F_{i_{j+1}}$,  with the
property
\begin{equation}\label{compensation}
\sum_{k=1}^j \frac{f_{i_k}}{a_{i_k}} < R \leq \sum_{k=1}^{j+1}
\frac{f_{i_k}}{a_{i_k}}
\end{equation}
we find a distribution of resource $\mathbf{r}_{\{{i_1},...
{i_{j+1}}\}}=(r_{i_1},... r_{i_{j+1}})$:
\begin{equation}\label{antiLiebigDistr}
r_{i_k}=\frac{f_{i_k}}{a_{i_k}} \ \ (k=1,... j)\ , \ \
r_{i_{j+1}}=R-\sum_{k=1}^j \frac{f_{i_k}}{a_{i_k}}\ , \ \ r_i=0 \
\ {\rm for }\ \ i \notin \{{i_1},... {i_{j+1}}\} \ .
\end{equation}
This distribution (\ref{compensation}) means that the pressure of
$j$ factors are completely compensated and one factor is partially
compensated. For $j=0$, Eq.~(\ref{compensation}) gives $0 < R \leq
f_{i_1}$ and there exists only one nonzero component in the
distribution (\ref{antiLiebigDistr}), $r_{i_{1}}=R$. For $j=q$ all
$r_i=f_i/a_i$, $\sum_i r_i < R$ and all factors are fully
compensated.

We get the following theorem as an application of standard results
about extreme points of convex sets \cite{Rockafellar} to the
monotonic function $W$ (\ref{monotonicity}) with strictly convex
sublevel sets.

\begin{theorem}\label{Theorem2} Any maximizer for $W(f_1-a_1r_1, ... f_q-a_q
r_q)$ under given conditions has the form $\mathbf{r}_{\{{i_1},...
{i_{j+1}}\}}$ (\ref{antiLiebigDistr}).
\end{theorem}

To find the optimal distribution we have to analyze which
distribution of the form (\ref{compensation}) gives the highest
fitness.

\begin{figure} \centering{
\includegraphics[width=100mm]{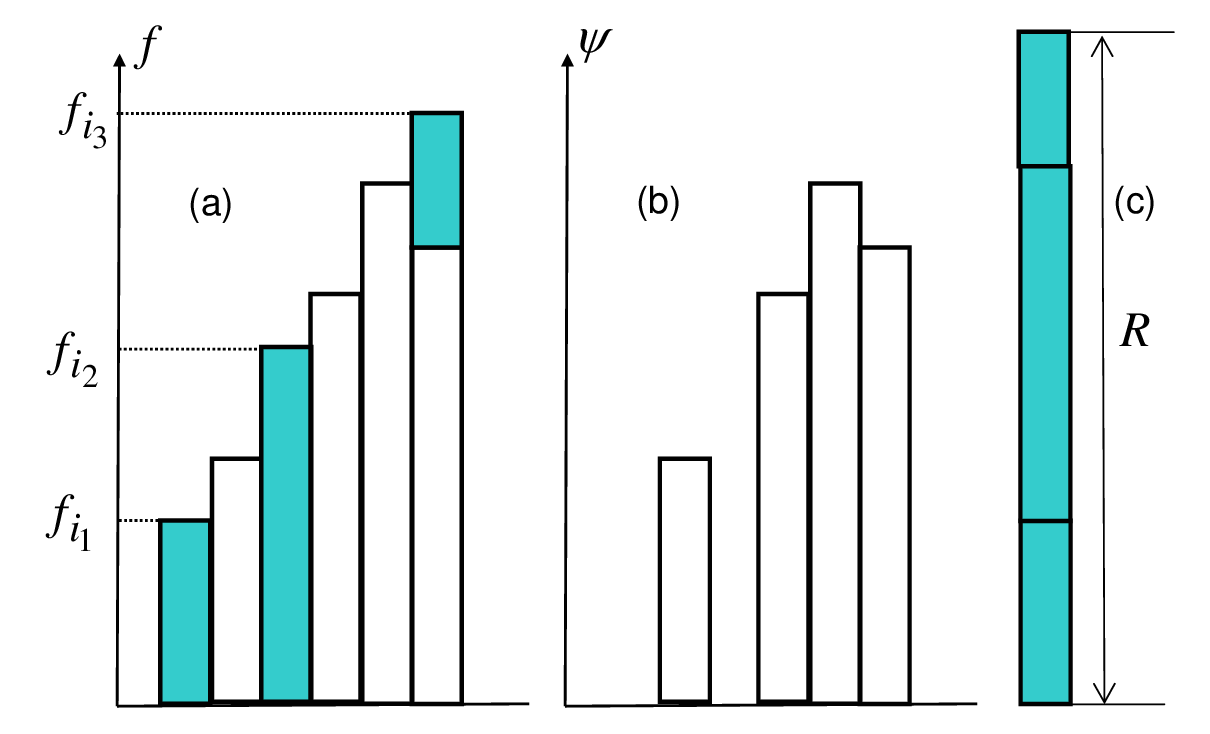}}
\caption{\label{Fig:FactorDistribSyn} Typical optimal distribution
of resource for neutralization of synergistic factors. (a) Factors
intensity (the compensated parts of factors are highlighted,
$j=2$), (b) distribution of tensions $\psi_i$ after adaptation
becomes less uniform (compare to Fig.~\ref{Fig:FactorDistrib}),
(c) the sum of distributed resources. For simplicity of the
picture, we take here all $a_i=1$.}
\end{figure}

If the initial distribution of factors intensities,
$\mathbf{f}=(f_1,... f_q)$, is almost uniform and all factors are
significant then, after adaptation, the distribution of effective
tensions, $\mathbf{\psi}=(\psi_1,... \psi_q)$ ($\psi_i=f_i-a_i
r_i$), is less uniform. Following Theorem~\ref{Theorem2}, some of
factors may be completely neutralized and one additional factor may
be neutralized partially. This situation is opposite to adaptation
to Liebig's system of factors, where amount of significant factors
increases and the distribution of tensions becomes more uniform
because of adaptation. For Liebig's system, adaptation transforms
low dimensional picture (one limiting factor) into high dimensional
one, and we expect the well-adapted systems have less correlations
than in stress. For synergistic systems, adaptation transforms high
dimensional picture into low dimensional one (less factors), and our
expectations are inverse: we expect the well-adapted systems have
more correlations than in stress (this situation is illustrated in
Fig.~\ref{Fig:FactorDistribSyn}; compare to
Fig.~\ref{Fig:FactorDistrib}). We call this property of adaptation
to synergistic system of factors the {\it Law of the Minimum inverse
paradox}.

The fitness by itself is a theoretical construction based on the
average reproduction coefficient (instant fitness). It is impossible
to measure this quantity in time intervals that are much shorter
than the life length and even for the life--long analysis it is a
non-trivial problem \cite{Shawatal2008}.

In order to understand which system of factors we deal with,
Liebig's or synergistic one, we have to compare theoretical
consequences of their properties and compare them to empirical data.
First of all, we can measure results of adaptation, and use  for
analysis properties of optimal adaptation in ensembles of systems
for analysis (Fig.~\ref{Fig:FactorDistrib},
Fig.~\ref{Fig:FactorDistribSyn}).

\section{Empirical data}

In many areas of practice, from physiology to economics, psychology,
and engineering we have to analyze the behavior of groups of many
similar systems, which are adapting to the same or similar
environment. Groups of humans in hard living conditions (Far North
city, polar expedition, or a hospital, for example), trees under
influence of anthropogenic air pollution, rats under poisoning,
banks in financial crisis, enterprizes in recession, and many other
situations of that type provide us with plenty of important
problems, problems of diagnostics and prediction.

For many such situations it was found that the correlations between
individual systems are better indicators than the value of
attributes. More specifically, in thousands of experiments it was
shown that in crisis, typically, even before obvious symptoms of
crisis appear, the correlations increase, and, at the same time, the
variance increase too. After the crisis achieves its bottom, it can
develop into two directions: recovering (both the correlations and
the variance decrease) or fatal catastrophe (the correlations
decrease, but the variance continue to increase).

In this Sec. we review several sets of empirical results which
demonstrate this effect. Now, after 21 years of studying this effect
\cite{GorSmiCorAd1st,Sedov},  we maintain that this property is
universal for groups of similar systems that are sustaining a stress
and have an adaptation ability. On the other hand, situations with
inverse behavior were predicted theoretically and found
experimentally \cite{Mansurov}. This makes the problem more
intriguing.

Below, to collect information about strong correlations between many
attributes in one indicator, we evaluate the non-diagonal part of
the correlation matrix and delete terms with values below a
threshold $\alpha$ from the sum:
\begin{equation}\label{lpweight}
 G = \sum_{j>k, \ |r_{jk}|> \alpha}
|r_{jk}|.
\end{equation}
This quantity $G$ is a weight of the {\it correlation graph}. The
vertices of this graph correspond to variables, and these vertices
are connected by edges, if the absolute value of the correspondent
sample correlation coefficient exceeds $\alpha$: $|r_{jk}|>\alpha$.
Usually, we take $\alpha=0.5$  (a half of the maximum) if there is
no reason to select another value.

\subsection{Adaptation of Adults for Change of Climatic Zone}

The activity of enzymes in human leukocytes was studied
\cite{Bul1Limf,Bul2Limf} (alkaline phosphatase, acid phosphatase,
succinate dehydrogenase, glyceraldehyde-3-phosphate dehydrogenase,
glycerol- 3-phosphate dehydrogenase, and glucose-6-phosphate
dehydrogenase).

\begin{figure} \centering{
\includegraphics[width=80mm]{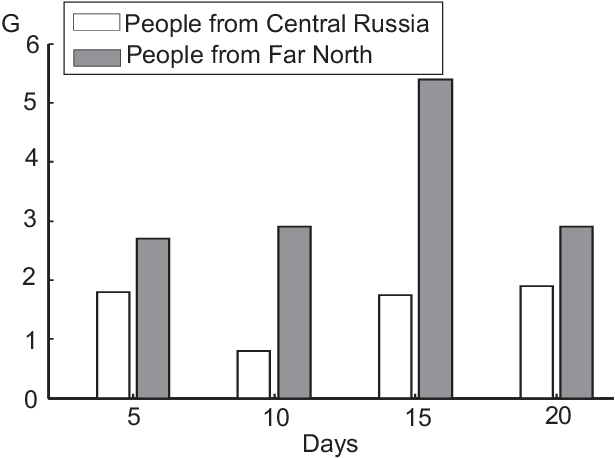}}
\caption{\label{Fig4:UrgentNorilsk}Weight of the correlation
graphs of activity of enzymes in leucocytes during urgent
adaptation at a resort. For people from Far North, the adaptation
crisis occurs near the 15th day.}
\end{figure}

We analyzed the short-term adaptation (20 days) of groups of healthy
20-30 year old men who change their climate zone:
\begin{itemize}
\item{From the Far North to the South resort (Sochi, Black Sea) in the summer;}
\item{From the temperate belt of Russia to the South resort (Sochi, Black Sea) in summer.}
\end{itemize}
Results are represented in Fig.~\ref{Fig4:UrgentNorilsk}. This
analysis supports the basic hypothesis and, on the other hand, could
be used for prediction of the most dangerous periods in adaptation,
which need special care.

We selected the group of 54 people who moved to the Far North, that
had any illness during the period of short-term adaptation. After 6
months in the Far North, this test group demonstrates much higher
correlations between activity of enzymes than the control group (98
people without illness during the adaptation period). For the
activity of enzymes in leucocytes $G=5.81$ in the test group versus
$G= 1.36$ in the control group. To compare the dimensionless
variance for these groups, we normalize the activity of enzymes to
unite sample means (it is senseless to use the trace of the
covariance matrix without normalization because normal activities of
enzymes differ in order of magnitude). For the test group, the sum
of the enzyme variances is  1.204, and for the control group it is
0.388.

\subsection{Collapse of Correlations ``on the Other Side of Crisis": Acute Hemolytic Anemia in Mice}

It is very important to understand where the system is going: (i)
to the bottom of the crisis with possibility to recover after that
bottom, (ii) to the normal state, from the bottom, or (iii) to the
``no return" point, after which it cannot recover.

This problem was studied in many situations with analysis of fatal
outcomes in oncological \cite{MansurOnco} and cardiological
\cite{Strygina} clinics, and also in special experiments with acute
hemolytic anemia caused by phenylhydrazine in mice \cite{mice}. The
main result here is: when approaching the no-return point,
correlations destroy ($G$ decreases), and variance typically does
continue to increase.

There exist no formal criterion to recognize the situation ``on the
other side of crisis". Nevertheless, it is necessary to select
situations for testing our hypothesis. Here the ``general
practitioner point of view" \cite{GP_AE1952} can be of help. From
such a point of view based on practical experience, the situation
described below is on the other side of crisis: the acute hemolytic
anemia caused by phenylhydrazine in mice with lethal outcome.

\begin{figure} \centering{
\includegraphics[width=60mm]{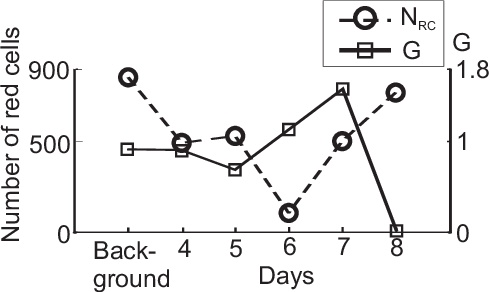}}
\caption{\label{Fig4:mice}Adaptation and disadaptation dynamics
for mice after phenylhydrazine injection.}
\end{figure}

This effect was demonstrated in special experiments \cite{mice}.
Acute hemolytic anemia caused by phenylhydrazine was studied in
CBAxlac mice.  Dynamics of correlation between hematocrit,
reticulocytes, erythrocytes, and leukocytes in blood is presented in
Fig.~\ref{Fig4:mice}. After phenylhydrazine injections (60 mg/kg,
twice a day, with interval 12 hours) during first 5-6 days the
amount of red cells decreased (Fig.~\ref{Fig4:mice}), but at the 7th
and 8th days this amount increased because of spleen activity. After
8 days most of the mice died. Weight of the correlation graph
increase precedeed the active adaptation response, but $G$ decreased
to zero before death (Fig.~\ref{Fig4:mice}), while amount of red
cells increased also at the last day.

\subsection{Grassy Plants Under Trampling Load}

\begin{table}\caption{Weight $G$ of the correlation graph for different grassy plants
under various trampling load  \label{GrassTrampl}} \centering{
\begin{tabular}{|l|c|c|c|} \hline
~Grassy Plant~ & ~Group 1~ & ~Group 2~ & ~Group 3~ \\
\hline
Lamiastrum   &     1.4   & 5.2  &  6.2  \\
 Paris (quadrifolia) & 4.1            & 7.6 & 14.8 \\
Convallaria   & 5.4    & 7.9               &  10.1  \\
Anemone   & 8.1                    & 12.5               &  15.8 \\
Pulmonaria   &  8.8        & 11.9     &  15.1   \\
Asarum & 10.3  & 15.4   & 19.5   \\
 \hline
\end{tabular}
}\end{table}

The effect exists for plants too. The grassy plants in oak
tree-plants are studied \cite{RazzhevaikinTrava1996}. For analysis
the fragments of forests are selected, where the densities of trees
and bushes were the same. The difference between those fragments was
in damaging the soil surface by trampling. Three groups of fragments
are studied:
\begin{itemize}
\item{Group 1 -- 0\% of soil surface are destroyed by trampling;}
\item{Group 2 -- 25\% of soil surface are destroyed by trampling;}
\item{Group 3 -- 70\% of soil surface are destroyed by trampling.}
\end{itemize}

The studied physiological attributes were: the height of sprouts,
the length of roots, the diameter of roots, the amount of roots, the
area of leafs, the area of roots. Results are presented in
Table~\ref{GrassTrampl}.

\subsection{Scots Pines Near a Coal Power Station}

The impact of emissions from a heat power station on Scots pine was
studied \cite{KofmantREES}. For diagnostic purposes the secondary
metabolites of phenolic nature were used. They are much more stable
than the primary products and hold the information about past impact
of environment on the plant organism for longer time.

The test group consisted of Scots pines (Pinus sylvestris L)  in a
40 year old stand of the II class in the emission tongue 10 km from
the power station. The station had been operating on brown coal for
45 years. The control group of Scots pines was from a stand of the
same age and forest type, growing outside the industrial emission
area. The needles for analysis were one year old from the shoots in
the middle part of the crown. The samples were taken in spring
during the bud swelling period. Individual composition of the
alcohol extract of needles was studied by high efficiency liquid
chromatography.  26 individual phenolic compounds were identified
for all samples and used in analysis.

No reliable difference was found in the test group and control
group average compositions. For example, the results for
Proantocyanidin content (mg/g dry weight) were as follows:
\begin{itemize}
\item{Total 37.4$\pm$3.2 (test) versus 36.8$\pm$2.0 (control);}
\end{itemize}
Nevertheless, the variance of compositions of individual compounds
in the test group was significantly higher, and the difference in
correlations was huge: $G=17.29$ for the test group versus
$G=3.79$ in the control group.

\subsection{Choice of Coordinates and the Problem of Invariance}

All indicators of the level of correlations are non-invariant with
respect to transformations of  coordinates. For example, rotation to
the principal axis annuls all the correlations. Dynamics of variance
also depends on nonlinear transformations of scales. Dimensionless
variance of logarithms (or ``relative variance") often demonstrates
more stable behavior especially when changes of mean values are
large. The observed effect depends on the choice of attributes.
Nevertheless, many researchers observed it without a special choice
of coordinate system. What does it mean? We can propose a
hypothesis: the effect may be so strong that it is almost improbable
to select a coordinate system where it vanishes. For example, if one
accepts the Selye model (\ref{1factorTension}), (\ref{OneFactor})
then observability of the effect means that for typical nonzero
values of $\psi$ in crisis
\begin{equation}\label{lrgeeffect}
l_k^2 \psi^2> {\rm var}(\epsilon_k)
\end{equation}
for more than one value of $k$, where var stands for variance of the
noise component (this is sufficient for increase of the
correlations). If $$\psi^2 \sum_k l_k^2 \gg \sum_k {\rm
var}(\epsilon_k)$$ and the set of allowable transformations of
coordinates is bounded (together with the set of inverse
transformations), then the probability to select randomly a
coordinate system which violates condition (\ref{lrgeeffect}) is
small (for reasonable definitions of this probability and of the
relation $\gg$).

\section{Comparison to Econometrics}

The simplest Selye's model (\ref{1factorTension}) seems very similar
to the classical one-factor econometrics models \cite{Campbell}
which assume that the returns of stocks ($\rho _i$) are controlled
by one factor, the ``market" return $M(t)$. In this model, for any
stock
\begin{equation}\label{OneFacEcon}
\rho _i(t)=a_i + b_i M(t)+ \epsilon_i(t)
\end{equation}
where $\rho_i(t)$  is the return of the $i$th stock at time $t$,
$a_i$ and $b_i$ are real parameters, and $\epsilon_i(t)$ is a zero
mean noise. In our models, the pressure of factor characterizes the
time window and is slower variable than the return.

The main difference between models (\ref{1factorTension}) and
(\ref{OneFacEcon}) could be found in the nonlinear coupling
(\ref{OneFactor}) between the environmental property (the factor
value $f$) and the property of individuals (the resource amount
$R$). Exactly this coupling causes separation of a population into
two groups: the well-adapted less correlated group and the highly
correlated group with larger variances of individual properties and
amount of resource which is not sufficient for compensation of the
factor load. Let us check whether such a separation is valid for
financial data.

\subsection{Data Description}

For the analysis of correlations in financial systems  we used the
daily closing values for companies that are registered in the FTSE
100 index (Financial Times Stock Exchange Index). The FTSE 100 is a
market-capitalization weighted index representing the performance of
the 100 largest UK-domiciled blue chip companies which pass
screening for size and liquidity. The index represents approximately
88.03\% of the UK's market capitalization.  FTSE 100 constituents
are all traded on the London Stock Exchange's SETS trading system.
We selected 30 companies that had the highest value of the capital
(on the 1st of January 2007) and  stand for different types of
business as well. The list of the companies and business types is
displayed in Table~\ref{CompList}.

\begin{table}\caption{Thirty largest companies for analysis
from the FTSE 100 index \label{CompList}} \centering{\small
\begin{tabular} {|c|l|l|l|}
  \hline
   Number &Business type & Company &  Abbreviation \\
  \hline \hline
  1& Mining & Anglo American plc& AAL\\
  2& & BHP Billiton& BHP\\
   \hline
  3& Energy (oil/gas) & BG Group & BG \\
  4& & BP & BP\\5&  & Royal Dutch Shell & RDSB \\
  \hline
  6 & Energy (distribution) & Centrica& CNA\\7& & National Grid & NG\\
  \hline
8&  Finance (bank) & Barclays plc & BARC\\9&   & HBOS & HBOS\\10& & HSBC HLDG & HSBC \\
 11& & Lloyds & LLOY\\
   \hline
  12& Finance (insurance)  & Admiral& ADM \\13& & Aviva & AV\\14& & LandSecurities&LAND\\
15&   & Prudential& PRU\\16& & Standard Chartered& STAN\\
    \hline
    17& Food production & Unilever &ULVR\\
    \hline
18& Consumer  & Diageo & DGE\\
19& goods/food/drinks & SABMiller& SAB\\
20& & TESCO &TSCO\\
\hline
21& Tobacco & British American Tobacco &BATS\\
22& & Imperial Tobacco & IMT\\
\hline
23& Pharmaceuticals& AstraZeneca &AZN\\
24& (inc. research)& GlaxoSmithKline & GSK\\
\hline
 25& Telecommunications & BT Group &BTA\\
 26& & Vodafone &VOD\\
\hline
27&Travel/leasure& Compass Group & CPG\\
\hline
28&Media (broadcasting) & British Sky Broadcasting & BSY\\
\hline
29& Aerospace/ & BAE System & BA\\
30& defence & Rolls-Royce& RR\\
\hline \hline
\end{tabular}
}\end{table}

\begin{figure}\centering{
a)\includegraphics[width=65mm]{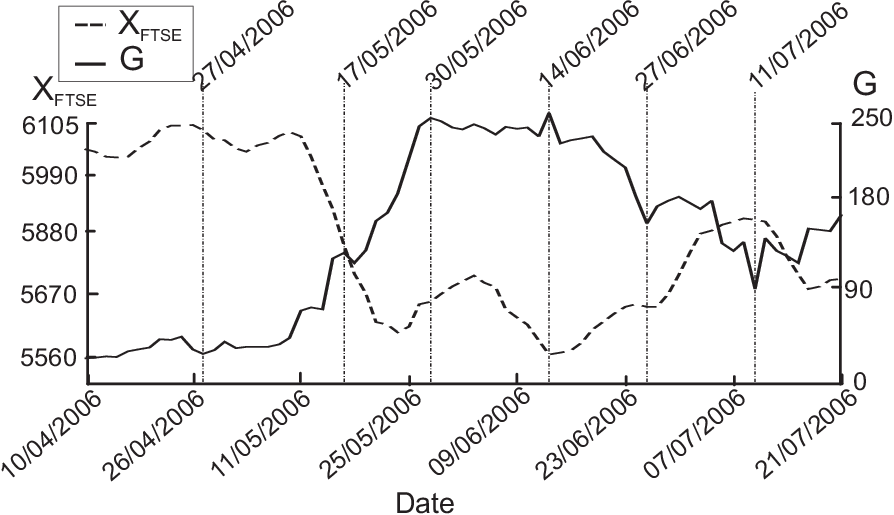}b)\includegraphics[width=40mm]{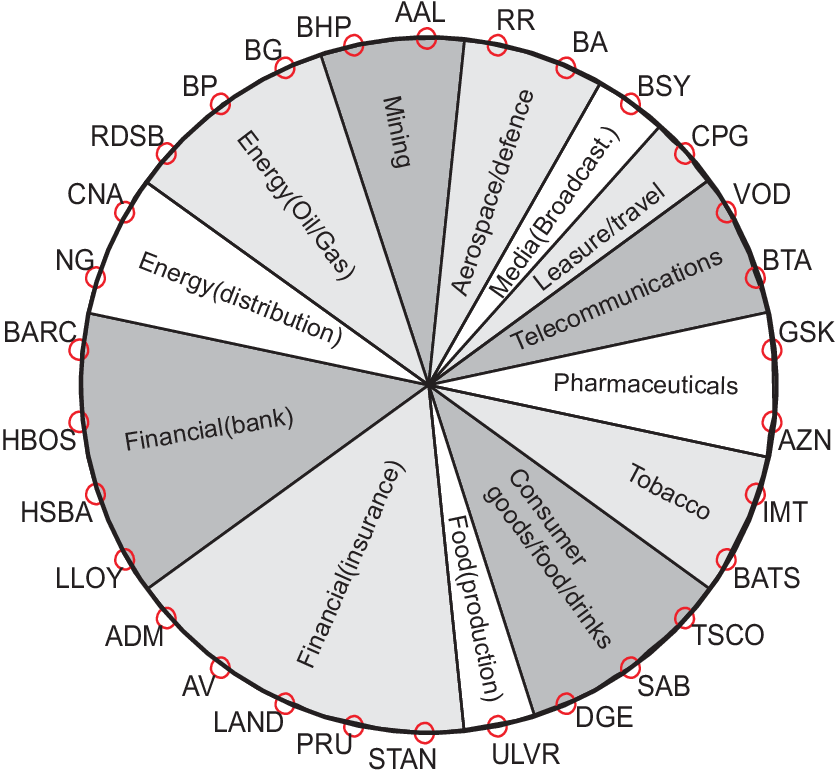}\\
 c)\includegraphics[width=105mm]{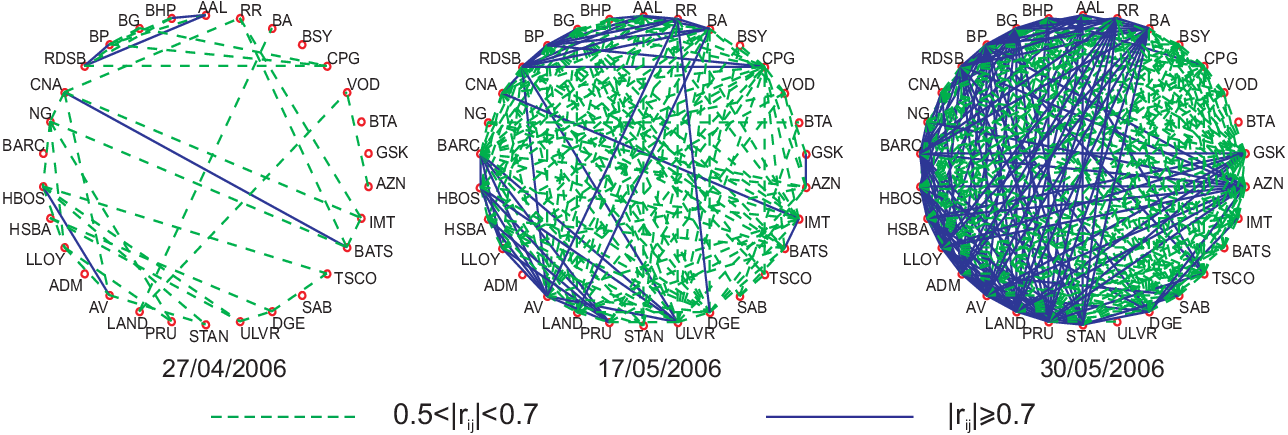} \\
 d)\includegraphics[width=105mm]{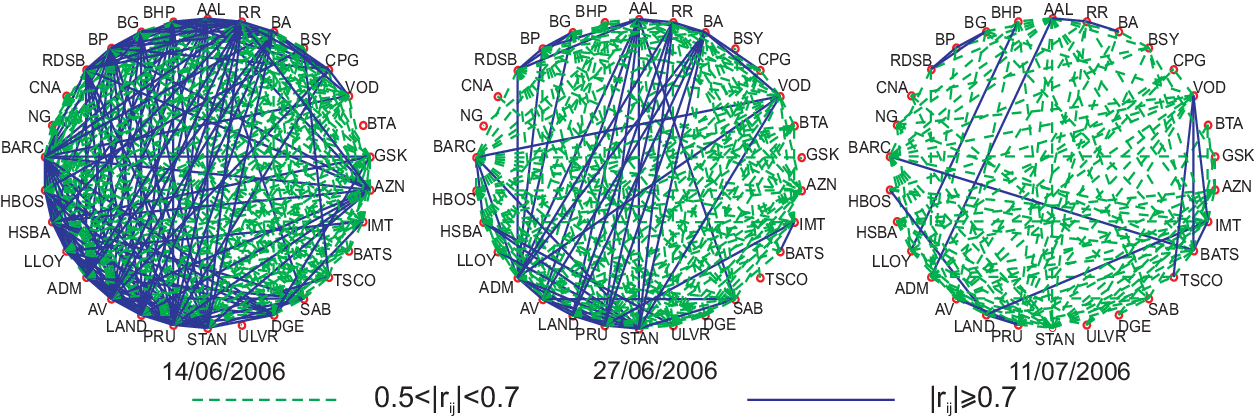}}
\caption{Correlation graphs for six positions of sliding time window
 on interval 10/04/2006 - 21/07/2006.
a)~Dynamics of FTSE100 (dashed line) and of $G$ (solid line) over
the interval, vertical lines correspond to the points that were used
for the correlation graphs.
 b)~Thirty companies for analysis and their distributions over
various sectors of economics.
 c)~The correlation graphs for the
first three points, FTSE100 decreases, the correlation graph becomes
more connective. d)~The correlation graphs for the last three
points, FTSE100 increases, the correlation graph becomes less
connective.\label{graph_int1}}
\end{figure}

Data for these companies are available form the Yahoo!Finance
web-site. For data cleaning we use also information for the selected
period available at the London Stock Exchange web-site. Let $x_i(t)$
denote the closing stock price for the $i$th company at the moment
$t$, where $i=\overline{1,30}$, $t$ is the discrete time (the number
of the trading day). We analyze the correlations of logarithmic
returns: $x^l _i(t)=\ln \frac{x_i (t)} {x_i(t-1)}$, in sliding time
windows of length $p=20$, this corresponds approximately to 4 weeks
of 5 trading days. The correlation coefficients $r_{ij}(t)$ for time
moment $t$ are calculated in the time window $[t-p,t-1]$, which
strongly precedes $t$. Here we calculate correlations between
individuals (stocks), and for biological data we calculated
correlations between attributes. This corresponds to transposed data
matrix.

\subsection{Who Belongs to the Highly Correlated Group in Crisis}

For analysis we selected the time interval 10/04/2006 - 21/07/2006
that represents the FTSE index decrease and restoration in spring
and summer 2006 (more data are analyzed in our e-print
\cite{GSTarXiv}). In Fig.~\ref{graph_int1} the correlation graphs
are presented for three time moments during the crisis development
and three moments of the restoration. The vertices of this graph
correspond to stocks. These vertices are connected by solid lines is
the correspondent correlation coefficient $|r_{jk}| \geq \sqrt{0.5}$
$(\sqrt{0.5}=\cos(\pi/4)\approx 0.707)$, and by dashed lines if
$\sqrt{0.5} > |r_{jk}|>0.5$.

The correlation graphs from Fig.~\ref{graph_int1} show that in the
development of this crisis (10/04/2006 - 21/07/2006) the correlated
group was formed mostly by two clusters: a financial cluster (banks
and insurance companies) and an energy (oil/gas) -- mining --
aerospace/defence and travel cluster. At the bottom of the crisis
the correlated phase included almost all stocks. The recovery
followed a significantly different trajectory: the correlated phase
in the recovery seems absolutely different from that phase in the
crisis development: there appeared the strong correlation between
financial sector and industry. This is a sign that after the crisis
bottom the simplest Selye's model is not valid for a financial
market. Perhaps, interaction between enterprizes and redistribution
of resource between them should be taken into account. We need
additional equations for dynamics of the available amounts of
resource  $R_i$ for $i$th stock. Nevertheless, appearance of the
highly correlated phase in the development of the crisis in the
financial world followed the predictions of Selye's model, at least,
qualitatively.

Asymmetry between the drawups and the drawdowns of the financial
market was noticed also in the analysis of the financial empirical
correlation matrix of the 30 companies which compose the Deutsche
Aktienindex (DAX) \cite{DrozdComCollNoise2000}.

The market mode was studied by principal component analysis
\cite{Stanley2002}. During periods of high market volatility values
of the largest eigenvalue of the correlation matrix are large.  This
fact was commented as a strong collective behavior in regimes of
high volatility. For this largest eigenvalue, the distribution of
coordinates of the correspondent eigenvector has very remarkable
properties:
\begin{itemize}
\item{It is much more uniform than the prediction of the random matrix
theory (authors of Ref.~\cite{Stanley2002} described this vector as
``approximately uniform", suggesting that all stocks participate in
this ``market mode");}
\item{Almost all components of that eigenvector have the same sign.}
\item{A large degree of cross correlations between stocks can
be attributed to the influence of the largest eigenvalue and its
corresponding eigenvector}
\end{itemize}
Two interpretations of this eigenvector were proposed
\cite{Stanley2002}: it corresponds either to the common strong
factor that affects all stocks, or it represents the ``collective
response" of the entire market to stimuli. Our observation supports
this conclusion at the bottom of the crisis. At the beginning of the
crisis  the correlated group includes stocks which are sensitive to
the factor load, and other stocks are tolerant and form the less
correlated group with the smaller variance. Following Selye's model
we can conclude that the effect is the result of nonlinear coupling
of the environmental factor load and the individual adaptation
response.

\section{Functional Decomposition and Integration  of Subsystems}

In the simple factor--resource Selye models the adaptation response
has no structure: the organism just distributes the adaptation
resource to neutralization of various harmful factors. It is
possible to make this model more realistic by decomposition. The
resource is assigned not directly ``against factors" but is used for
activation and intensification of some subsystems.

We need to define the hierarchical structure of the organism to link
the  behavior in across multiple scales. In integrative and
computational physiology it is necessary to go both bottom--up and
top--up approaches. The bottom--up approach goes from proteins to
cells, tissues, organs and organ systems, and finally to a whole
organism \cite{Cramlinatal2004}.

The top-down approach starts from a bird's eye view of the behavior
of the system -- from the top or the whole and aims to discover and
characterize biological mechanisms closer to the bottom -- that is,
the parts and their interactions \cite{Cramlinatal2004}.

There is a long history of discussion of functional structure of the
organism, and many approaches are developed: from the Anokhin theory
of functional systems \cite{Sudakov2004} to the inspired by the
General Systems approach theory of ``Formal Biological Systems"
\cite{Chauvet1999}.

The notion of functional systems represents a special type of
integration of physiological functions. Individual organs and tissue
elements, are selectively combined into self-regulating systems
organizations to achieve the necessary adaptive results important
for the whole organism. The self-organization process is ruled by
the adaptation needs.

For decomposition of the models of physiological systems, the
concept of principal dynamic modes was developed
\cite{Marmarelis1997,Marmarelis2004}.

In this section, we demonstrate how to decompose the
factor--resource models of the adaptation of the organism to
subsystems.

In general, the analysis of interaction of factors is decomposed to
interaction of factors and subsystems. Compensation of the harm from
each factor $F_i$ requires activity of various systems. For every
system $S_j$ a variable, activation level $I_j$ is defined. Level 0
corresponds to a fully disabled subsystem (and for most of
essentially important subsystems it implies death). For each factor
$F_i$ and every subsystem $S_j$ a ``standard level" of activity
$\varsigma_{ij}$ is defined. Roughly speaking, this level of
activation of the subsystem $S_j$ is necessary for neutralization of
the unit value of the pressure of the factor $F_i$. If
$\varsigma_{ij}=0$ then the subsystem $S_j$ is not involved in the
neutralization of the factor $F_i$.

Instead of $\psi=f-ar_f$ (see (\ref{OneFactor})) we have to use
$$\omega_i=f_i-\min_{j, \ \varsigma_{ij}\neq
0}\left\{\frac{I_j}{\varsigma_{ij}}\right\}\ , $$ and the
compensated value of the factor pressure $F_i$ is
\begin{equation}\label{decomposition}
\psi_i=
 \left\{\begin{array}{ll}
  &\omega_i\ , \ \ {\rm if} \ \ \omega_i>0 \ ; \\
 &0 ,\ \  \ {\rm else \ .}
  \end{array}
 \right.
\end{equation}

In this model resources are assigned not to neutralization of
factors but for activation of subsystems. The activation intensity
of the subsystem $S_j$ depends on the adaptation resource $r_j$,
assigned to this subsystem:
\begin{equation}\label{activationassignation}
I_j=\alpha_j r_j \ \ {\rm and} \ \  \omega_i=f_i-\min_{j, \
\varsigma_{ij}\neq 0}\left\{\frac{\alpha_j}{\varsigma_{ij}}r_j
\right\}\ .
\end{equation}

For any given organization of the system of factors, optimization of
fitness together with definitions (\ref{decomposition}) and
(\ref{activationassignation}) lead to a clearly stated optimization
problem. For example, for Liebig's system of factors we have to find
distributions of $r_j$ that give solution to a problem:
\begin{equation}\label{liebDecompoz1}
\max_i \psi_i \to \min  \ \ {\rm for } \ \ r_j\geq 0, \ \sum_j r_j
\leq R \ .
\end{equation}
If this minimum of maxima is positive ($\min (\max_i \psi_i)>0$)
then the optimal distribution of resources is unique. If $\min
(\max_i \psi_i)=0$ then there exists a polyhedron of optimal
distributions given by the system of inequalities:

$$\frac{\alpha_j}{\varsigma_{ij}}r_j \geq f_i \ , \
\ r_j\geq 0  \ \ \mbox{for all} \ \ i,j, \ \ \sum_j r_j \leq R \ .$$

For the study of integration in experiment we use principal
component analysis and find, parameters of which systems give
significant inputs in the first principal components. Under the
stress, the configuration of the  subsystems, which are
significantly involved in the first principal components, changes
\cite{Svetlichnaia1997}.

We analyzed interaction of cardiovascular and respiratory subsystems
under exercise tolerance tests at various levels of load. Typically,
we observe the following dynamics of the first factor composition.
With increase of the load, coordinates both the correlations of the
subsystems attributes with the first factor increase up to some
maximal load which depend on the age and the health in the group of
patients. After this maximum of integration, if the load continues
to increase then the level of integration decreases
\cite{Svetlichnaia1997}.

Generalization of Selye's models by decomposition creates a rich and
flexible system of models for adaptation of hierarchically organized
systems. Principal component analysis \cite{Jolliffe2002} with its
various nonlinear generalizations
\cite{Gorbanatal2008,GorbanZinovyev2009} gives a system of tools for
extracting the information about integration of subsystems from the
empirical data.

\section{Conclusion}

Due to the Law of the Minimum paradoxes, if we observe the Law of
the Minimum in artificial systems, then under natural conditions
adaptation will equalize the load of different factors and we can
expect a violation of the Law of the Minimum. Inversely, if an
artificial systems demonstrate significant violation of the law of
the minimum, then we can expect that under natural conditions
adaptation will compensate this violation.

This effect follows from the factor--resource models of adaptation
and the idea of optimality applied to these models. We don't need an
explicit form of generalized fitness (which may be difficult to
find), but use only the general properties that follow from the Law
of the Minimum (or, oppositely, from the assumption of synergy).

Another consequence of the factor--resource models is the prediction
of the appearance of strongly correlated groups of individuals under
an increase of the load of environmental factors. Higher
correlations in those groups do not mean that individuals become
more similar, because the variance in those groups is also higher.
This effect is observed for financial market too and seems to be
very general in ensembles of systems which are adapting to
environmental factors load.

Decomposition of the factor--resource models for the hierarchy of
subsystems allows us to discuss integration of the subsystems in
adaptation. For the explorative analysis of this integration in
empirical data the principal component analysis is the first choice:
for the high level of integration different subsystems join in the
main factors.

The most important  shortcoming of the factor--resource models is
the lack of dynamics. In the present form it describes adaptation as
a single action, the distribution of the adaptation resource. We
avoid any kinetic modeling. Nevertheless, adaptation is a process in
time. We have to create a system of dynamical models.


\begin{thebibliography}{99}

\bibitem{Liebig1}F. Salisbury (1992)
Plant physiology (4th ed.), Wadsworth, Belmont, CA.

\bibitem{van der Ploeg1999}R.R. van der Ploeg, W. B\"ohm, M.B.
Kirkham, (1999), On the origin of the theory of mineral nutrition of
plants and the law of the minimum, Soil Science Society of America
Journal 63, 1055--1062.

\bibitem{Liebig2+}Q. Paris, (1992), The Return of von Liebig's ``Law of the Minimum", Agron.
J., 84, 1040--1046

\bibitem{Liebig3+}B.S. Cade, J.W. Terrell, R.L. Schroeder, (1999), Estimating effects of
limiting factors with regression quantiles, Ecology 80 (1),
311--323.

\bibitem{Tilman1980}D. Tilman, (1980), Resources: a graphical-mechanistic approach to
competition and predation, Am. Nat. 116 (3), 362--393.

\bibitem{Tilman1982}D. Tilman  (1982) Resource Competition and Community Structure, Princeton
University Press, Princeton, NJ.

\bibitem{BloomChapinMooney1985}A.J. Bloom, F.S. Chapin III,
H.A. Mooney,  (1985), Resource limitation in plants -- an economic
analogy. Annu. Rev. Ecol. Syst. 16, 363--392.

\bibitem{ChapinSchulzeMooney1990}
F.S. Chapin, III, E. Schulze, H.A. Mooney, (1990), The Ecology and
Economics of Storage in Plants, Annu. Rev. Ecol. Syst.  21,
423--447.

\bibitem{GorSmiCorAd1st}A.N. Gorban, V.T. Manchuk, E.V.
Petushkova (Smirnova) (1987) Dynamics of physiological paramethers
correlations and the ecological-evolutionary principle of
polyfactoriality, In: Problemy Ekologicheskogo Monitoringa i
Modelirovaniya Ekosistem [The Problems of Ecological Monitoring and
Ecosystem Modelling], Vol. 10. Gidrometeoizdat, Leningrad, pp.
187--198.

\bibitem{Sedov}K.R. Sedov, A.N. Gorban', E.V. Petushkova (Smirnova), V.T. Manchuk, E.N.
Shalamova, (1988), Correlation adaptometry as a method of screening
of the population, Vestn Akad Med Nauk SSSR (10), 69--75. PMID:
3223045

\bibitem{SihGleeson1995}A. Sih, S.K. Gleeson, (1995), A limits-oriented approach to
evolutionary ecology, Trends in Ecology and Evolution 10 (9),
 378--382.

\bibitem{Kobe1996}R.K. Kobe (1996) Intraspecific Variation in Sapling Mortality and
Growth Predicts Geographic Variation in Forest Composition, In:
Ecological Monographs: Vol. 66  (2), pp. 181--201.

\bibitem{vandenBerg1998}H.A. van den Berg, (1998), Multiple nutrient limitation in
unicellulars: reconstructing Liebig's law, Mathematical Biosciences,
149 (1), 1--22.

\bibitem{Aumontatal2003}O. Aumont, E. Maier-Reimer, S. Blain, P. Monfray, (2003), An
ecosystem model of the global ocean including Fe, Si, P
colimitations, Global Biogeochemical Cycles 17 (2), 1060,
doi:10.1029/2001GB001745.

\bibitem{EgliZinn2003}T. Egli, M. Zinn, (2003), The concept of
multiple-nutrient-limited growth of microorganisms and its
application in biotechnological processes, Biotechnology Advances 22
(1-2), 35--43

\bibitem{ZinnWitholtEgli2004}M. Zinn, B. Witholt, T. Egli, (2004), Dual nutrient limited
growth: models, experimental observations, and applications, Journal
of Biotechnology 113 (1-3), 263--279.

\bibitem{WutzlerReichstein2008}T. Wutzler and M. Reichstein, (2008), Colimitation of decomposition
by substrate and decomposers -- a comparison of model formulations,
Biogeosciences Discuss. 5, 163--190.

\bibitem{Dangeratal2008}M. Danger, T. Daufresne, F. Lucas, S. Pissard, G. Lacroix,
(2008), Does Liebig's law of the minimum scale up from species to
communities? Oikos 117 (11), 1741--1751

\bibitem{SemSem}F.N.~Semevsky,  S.M.~Semenov (1982) Mathematical modeling of ecological
processes, Gidrometeoizdat, Leningrad.

\bibitem{SaitoGoepfert2008}M.A. Saito, T.J. Goepfert, (2008), Some thoughts on the concept
of colimitation: Three definitions and the importance of
bioavailability, Limnol. Oceanogr. 53 (1), 276--290

\bibitem{Nijlandatal2008}G.O. Nijland, J. Schouls, J. Goudriaan, (2008), Integrating the
production functions of Liebig, Michaelis-Menten, Mitscherlich and
Liebscher into one system dynamics model, NJAS - Wageningen Journal
of Life Sciences 55 (2), 199--224.

\bibitem{Brown}G. C. Brown and C. E. Cooper,  (1993). Control analysis applied
to a single enzymes: can an isolated enzyme have a unique
rate--limiting step? Biochem. J. 294, 87--94.

\bibitem{GorRadLim}A.N. Gorban, O. Radulescu, (2008),
Dynamic and Static Limitation in Multiscale Reaction Networks,
Revisited, Advances in Chemical Engineering 34, 103--173.

\bibitem{Droop1973}M.R. Droop, (1973), Some thoughts on nutrient limitation in
algae, J. Phycol. 9, 264--272.

\bibitem{LegovicCruzado1997}T. Legovi\'c, A. Cruzado, (1997), A model of phytoplankton growth
on multiple nutrients based on the Michaelis-Menten-Monod uptake,
Droop's growth and  Liebig's  law, Ecol. Modelling 99 (1), 19--31.

\bibitem{Ballantyneatal2008}F. Ballantyne IV., D.N.L. Menge, A. Ostling and P. Hosseini, (2008),
Nutrient recycling affects autotroph and ecosystem stoichiometry,
Am. Nat. 171 (4), 511--523.

\bibitem{Shoreshatal2008}N. Shoresh, M. Hegreness, R. Kishony, (2008), Evolution exacerbates the
paradox of the plankton, PNAS USA 105 (34), 12365--12369.

\bibitem{Hutchinson1961}G.E. Hutchinson, (1961), The paradox of the plankton,  Am. Nat.
95, 137--145.

\bibitem{Menge2009}D.N.L. Menge, J.S. Weitz, (2009), Dangerous nutrients: Evolution of
phytoplankton resource uptake subject to virus attack, Journal of
Theoretical Biology, 257 (1), 104--115.

\bibitem{Chertovatal2004}O. Chertov, A. Gorbushina, B. Deventer, (2004), A model for
microcolonial fungi growth on rock surfaces, Ecological Modelling
177 (3-4), 415--426.

\bibitem{McGill2005}B. McGill, (2005), A mechanistic model of a mutualism and its
ecological and evolutionary dynamics, Ecological Modelling 187 (4),
Pages 413--425.

\bibitem{Hennessy2009}D.A. Hennessy, (2009), Crop yield skewness under law of the minimum
technology, American Journal of Agricultural Economics 91 (1),
197--208.

\bibitem{Austin2007}M. Austin, (2007), Species distribution models and ecological theory: A
critical assessment and some possible new approaches, Ecological
Modelling 200 (1-2), 1--19.

\bibitem{Thomsonatal1996}J.D. Thomson, G. Weiblen, B.A. Thomson, S. Alfaro, P.
Legendre, (1996), Untangling multiple factors in spatial
distributions: lilies, gophers and rocks, Ecology 77, 1698–1715.

\bibitem{EcolEcon}H.E. Daly,  (1991), Population and Economics -- A Bioeconomic
Analysis,  Population and Environment 12 (3), 257--263.

\bibitem{EcolEdu}M.Y. Ozden, (2004), Law of the Minimum in Learning, Educational
Technology \& Society 7 (3), 5--8.

\bibitem{KoMa97}V. Kolokoltsov, V. Maslov (1997) Idempotent
analysis and applications, Kluwer Acad. Publ., Dordrecht.

\bibitem{LitvinovMaslov2005}G.L. Litvinov, V.P. Maslov (Eds.) (2005) Idempotent mathematics
and mathematical physics, Contemporary Mathematics, Vol. 377, AMS,
Providence, RI.

\bibitem{Litvinov2007}G.L. Litvinov, (2007), The Maslov dequantization, idempotent and
tropical mathematics: a brief introduction, Journal of Mathematical
Sciences 140 (3), 426-444. E-print: arXiv:math/0507014 [math.GM].

\bibitem{Kleene}S.C.~Kleene (1956) Representation of events in nerve sets and
finite automata, In: J.~McCarthy and C.~Shannon (Eds), Automata
Studies, Princeton University Press, Princeton, pp.~3--40.


\bibitem{GorbanRadZin2010}A.N. Gorban, O. Radulescu, A.Y. Zinovyev, (2010), Asymptotology of
chemical reaction networks, Chem. Eng. Sci. 65, 2310--2324.

\bibitem{Fisher1930}R.A. Fisher  (1930) The genetical theory of natural selection,
Oxford University Press, Oxford.

\bibitem{Haldane1932}J.B.S. Haldane  (1932) The causes of evolution, Longmans Green,
London.

\bibitem{MetzNisbetGeritz1992}J.A.J. Metz, R.M. Nisbet, S.A.H. Geritz,  (1992), How should we
define fitness for general ecological scenarios, Trends Ecol. Evol.
7, 198--202.

\bibitem{G1984}A.N. Gorban (1984) Equilibrium encircling. Equations of chemical kinetics and their
thermodynamic analysis, Nauka, Novosibirsk.

\bibitem{Maynard-Smith1982}J. Maynard-Smith (1982) Evolution and the
Theory of Games, Cambridge University Press, Cambridge.

\bibitem{WaxmanWelch2005}D. Waxman, J.J. Welch, (2005), Fisher's Microscope and Haldane's
Ellipse, Am. Nat. 166,  447--457.

\bibitem{KingsolverPfennig2007}J.G. Kingsolver, D.W. Pfennig, (2007), Patterns and Power of
Phenotypic Selection in Nature, BioScience 57 (7), 561--572.

\bibitem{Shawatal2008}R.G. Shaw, C.J. Geyer, S. Wagenius, H.H. Hangelbroek, J.R. Etterson,
(2008), Unifying Life-History Analyses for Inference of Fitness and
Population Growth, Am. Nat. 172, E35--E47.


\bibitem{Colborn}T. Colborn, D. Dumanoski, J.P. Meyers (1996) Our
Stolen Future: Are We Threatening Our Fertility, Intelligence, and
Survival? -- A Scientific Detective Story, Dutton, Peguin Books, NY.

\bibitem{SelyeAEN}H. Selye, (1938), Adaptation Energy, Nature 141 (3577), 926.

\bibitem{SelyeAE1}H. Selye, (1938), Experimental evidence supporting the conception of
``adaptation energy", Am. J. Physiol. 123, 758--765.

\bibitem{GP_AE1952}B. Goldstone, (1952), The general practitioner and the general adaptation
syndrome, S. Afr. Med. J.  26, 88--92, 106--109  PMID: 14901129,
14913266.

\bibitem{AEencicl}R. McCarty, K. Pasak  (2000) Alarm phase and general adaptation
syndrome, in: Encyclopedia of Stress, George Fink (ed.), Vol. 1,
Academic Press, pp. 126--130.

\bibitem{BreznitzAEappl}S. Breznitz (Ed.) (1983) The Denial of Stress, International
Universities Press, Inc., New York.

\bibitem{SchkadeOccAdAE2003}J.K. Schkade, S. Schultz (2003)  Occupational Adaptation in
Perspectives, Ch. 7 in: Perspectives in Human Occupation:
Participation in Life, By Paula Kramer, Jim Hinojosa, Charlotte
Brasic Royeen (eds), Lippincott Williams \& Wilkins, Baltimore, MD,
pp. 181--221.

\bibitem{Gause}G.F. Gause (1934)  The struggle for existence, Williams and Wilkins, Baltimore.
Online: http://www.ggause.com/Contgau.htm.

\bibitem{GorSlowRelax}A.N. Gorban (2004) Singularities of Transition Processes in Dynamical Systems: Qualitative Theory of Critical
Delays. Electron. J. Diff. Eqns., Monograph 05. E-print:
arXiv:chao-dyn/9703010.

\bibitem{Bom02}I.M. Bomze, (2002), {  Regularity vs. degeneracy in dynamics, games, and
optimization: a unified approach to different aspects.} SIAM Review
44, 394--414.

\bibitem{Oechssler02}J.~Oechssler,  F.~Riedel, (2002), {  On the Dynamic Foundation of
Evolutionary Stability in Continuous Models.} Journal of Economic
Theory  107, 223--252.

\bibitem{GorbanSelTth}A.N. Gorban, (2007), Selection Theorem for Systems with Inheritance, Math.
Model. Nat. Phenom. 2 (4),  1--45. E-print: arXiv:cond-mat/0405451
[cond-mat.stat-mech].

\bibitem{ZuckerkandlConvergGenEnv}E. Zuckerkandl,  R. Villet, (1988), Concentration-affinity
equivalence in gene regulation: Convergence of genetic and
environmental effects. PNAS USA, 85,  4784--4788.

\bibitem{West-Eberhardgenocopy-phenocopy}M.J. West-Eberhard (2003) Developmental Plasticity and Evolution,
Oxford University Press, US.

\bibitem{Hoffman1978}R.J. Hoffman, (1978), Environmental Uncertainty and Evolution of
Physiological Adaptation in Colias Butterflies,  Am. Nat. 112 (988),
999--1015.

\bibitem{Greene1999}E. Greene (1999) Phenotypic variation in larval development and
evolution: polymorphism, polyphenism, and developmental reaction
norms, In: The origin and evolution of larval forms, M. Wake, B.
Hall (Eds.), Academic Press, New York,  pp. 379–410.

\bibitem{Lerner1999}H.R. Lerner (ed.) (1999) Plant responses to environmental stresses:
from phytohormones to genome reorganization, Marcel Dekker, Inc.,
New York.

\bibitem{Miloatal2007}R. Milo, J.H. Hou, M. Springer, M.P. Brenner, M.W. Kirschner,
(2007), The relationship between evolutionary and physiological
variation in hemoglobin, PNAS USA 104 (43), 16998--17003.

\bibitem{FuscoMinelli2010}G. Fusco, A. Minelli, (2010), Phenotypic plasticity in development
and evolution: facts and concepts, Phil. Trans. R. Soc. B 365
(1540), 547--556.



\bibitem{Odum}E.P. Odum,  (1971), Fundamentals of ecology (Third Edition), W. B. Saunders,
Comp., Philadelphia -- London -- Toronto.

\bibitem{Rockafellar}R.T. Rockafellar, (1997), Convex analysis, Princeton University Press,
Princeton, NJ.

\bibitem{Mansurov}A.S. Mansurov, T.P. Mansurova, E.V. Smirnova, L.S. Mikitin,
A.V. Pershin (1994) How do correlations between physiological
parameters depend on the influence of different systems of stress
factors? In: Global \& Regional Ecological Problems, R.G. Khlebopros
(Ed.), Krasnoyarsk State Technical University Publ., Krasnoyarsk,
pp. 499--516.

\bibitem{Bul1Limf}G.V. Bulygin, A.S. Mansurov, T.P. Mansurova, E.V. Smirnova (1992) Dynamics
of parameters of human metabolic system during the short-term
adaptation, Institute of Biophysics, Russian Academy of Sciences,
Preprint 180B, Krasnoyarsk.

\bibitem{Bul2Limf}G.V. Bulygin, A.S. Mansurov, T.P. Mansurova, A.A. Mashanov, E.V.
Smirnova (1992) Impact of health on the ecological stress dynamics.
Institute of Biophysics, Russian Academy of Sciences, Preprint 185B,
Krasnoyarsk.

\bibitem{MansurOnco}A.S. Mansurov, T.P. Mansurova, E.V. Smirnova, L.S. Mikitin, A.V.
Pershin (1995) Human adaptation under influence of synergic system
of factors (treatment of oncological patients after operation),
Institute of Biophysics Russian Academy of Sciences, Preprint 212B
Krasnoyarsk.

\bibitem{Strygina}S.O. Strygina, S.N. Dement'ev, V.M. Uskov, G.I. Chernyshova (2000)
Dynamics of the system of correlations between physiological
parameters in patients after myocardial infarction, In: Mathematics,
Computer, Education, Proceedings of conference, Issue 7, Moscow,
 pp. 685--689.

\bibitem{mice}L.D. Ponomarenko, E.V. Smirnova, (1998), Dynamical characteristics of blood system
in mice with phenilhydrazin anemiya, In: Proceeding of 9th
International Symposium ``Reconstruction of homeostasis",
Krasnoyarsk, Russia, March 15-20, vol. 1, pp. 42--45.

\bibitem{RazzhevaikinTrava1996}I.V. Karmanova, V.N. Razzhevaikin, M.I.
Shpitonkov, (1996), Application of correlation adaptometry for
estimating a response of herbaceous species to stress loadings,
Doklady Botanical Sciences, Vols. 346--348, 4--7. [Translated from
Doklady Akademii Nauk SSSR, 346, 1996.]

\bibitem{KofmantREES}P.G. Shumeiko, V.I. Osipov, G.B. Kofman, (1994), Early detection of
industrial emission impact on Scots Pine needles by composition of
phenolic compounds, In: Global \& Regional Ecological Problems, R.G.
Khlebopros (Ed.), Krasnoyarsk State Technical University Publ.,
Krasnoyarsk, pp. 536--543.

\bibitem{Campbell}J.Y. Campbell, A.-W. Lo, and A.C. MacKinlay (1997) The Econometrics of
Financial Markets, Princeton Unviersity Press, Princeton.

\bibitem{GSTarXiv}A.N. Gorban, E.V. Smirnova, T.A. Tyukina, (2009),  Correlations, risk
and crisis: from physiology to  finance, E-print: arXiv:0905.0129
[physics.bio-ph].

\bibitem{DrozdComCollNoise2000}S. Dro\.{z}d\.{z}, F. Gr\"ummer , A.Z. G\'orski, F. Ruf, J. Speth,
(2000), Dynamics of competition between collectivity and noise in
the stock market, Physica A 287, 440--449.

\bibitem{Stanley2002}V. Plerou, P. Gopikrishnan, B. Rosenow, L.A.N. Amaral, T. Guhr,
H.E. Stanley, (2002), Random matrix approach to cross correlations
in financial data, Phys. Rev. E 65, 066126.

\bibitem{Cramlinatal2004}E.J. Crampin, M. Halstead, P. Hunter, P. Nielsen, D. Noble, N.
Smith, M. Tawhai, (2004), Computational physiology and the physiome
project, Experimental Physiology 89, 1-26.

\bibitem{Sudakov2004}K.V. Sudakov, (2004), Functional Systems Theory: A New Approach to
the Question of the Integration of Physiological Processes in the
Body, Neuroscience and Behavioral Physiology 34(5), 495--500.

\bibitem{Chauvet1999}G.A. Chauvet, (1999), S-propagators: a formalism for the
hierarchical organization of physiological systems. Application to
the nervous and the respiratory systems, International Journal of
General Systems 28 (1), 53--96.

\bibitem{Marmarelis1997}V.Z. Marmarelis, (1997), Modeling methodology for nonlinear
physiological systems, Annals of Biomedical Engineering  25, (2),
239--251.

\bibitem{Marmarelis2004}V.Z. Marmarelis (2004) Nonlinear Dynamic Modeling of Physiological
Systems, Wiley, New York.

\bibitem{Svetlichnaia1997}G.N. Svetlichnaia, E.V. Smirnova, L.I. Pokidysheva, (1997), Correlational
adaptometry as a method for evaluating cardiovascular and
respiratory interaction, Fiziol. Cheloveka 23 (3),  58--62. PMID:
9264951.

\bibitem{Jolliffe2002}I.T. Jolliffe  (2002) Principal component analysis, series:
Springer series in statistics, 2nd ed., XXIX, Springer, New York.

\bibitem{Gorbanatal2008}A.N. Gorban, B. Kegl, D. Wunsch, A. Zinovyev  (Eds.) (2008)
Principal Manifolds for Data Visualisation and Dimension Reduction,
Lecture Notes in Computational Science and Engineering, Vol. 58,
Springer, Berlin -- Heidelberg -- New York.

\bibitem{GorbanZinovyev2009}A.N. Gorban, A.Y. Zinovyev (2009) Principal Graphs and Manifolds,
Ch. 2 in: Handbook of Research on Machine Learning Applications and
Trends: Algorithms, Methods, and Techniques, Emilio Soria Olivas et
al. (eds), IGI Global, Hershey, PA, pp. 28--59. E-print:
arXiv:0809.0490 [cs.LG].

\end{thebibliography}
\end{document}